\documentclass[twocolumn,aps,superscriptaddress,showpacs]{revtex4}
\usepackage{hyperref}
\usepackage{amssymb}
\usepackage{amsmath}
\usepackage{graphicx}
\usepackage[normalem]{ulem}
\usepackage{url}
\usepackage{color}

\begin{document}
\title{
Comparison of heavy-ion transport simulations: Collision integral in a box\\
}

\author{Ying-Xun Zhang}
\email{zhyx@ciae.ac.cn}
\affiliation{China Institute of Atomic Energy, Beijing 102413, China}
\affiliation{Guangxi Key Laboratory Breeding Base of Nuclear Physics and Technology, Guilin 541004, China}

\author{Yong-Jia Wang}
\email{wangyongjia@zjhu.edu.cn}
\affiliation{School of Science, Huzhou University, Huzhou 313000,
China}

\author{Maria Colonna}
\email{colonna@lns.infn.it}
\affiliation{INFN-LNS, Laboratori Nazionali del Sud, 95123 Catania,
Italy}

\author{Pawel Danielewicz}
\email{danielewicz@nscl.msu.edu}
\affiliation{National Superconducting Cyclotron Laboratory and
Department of Physics and Astronomy, Michigan State
University, East Lansing, Michigan 48824, USA}

\author{Akira Ono}
\email{ono@nucl.phys.tohoku.ac.jp}
\affiliation{Department of Physics, Tohoku University, Sendai
980-8578, Japan}

\author{Manyee Betty Tsang}
\email{tsang@nscl.msu.edu}
\affiliation{National Superconducting Cyclotron Laboratory and
Department of Physics and Astronomy, Michigan State
University, East Lansing, Michigan 48824, USA}

\author{Hermann Wolter}
\email{hermann.wolter@physik.uni-muenchen.de}
\affiliation{Physics Department, University of Munich, D-85748 Garching, Germany}

\author{Jun Xu}
\email{xujun@sinap.ac.cn}
\affiliation{Shanghai Institute of Applied Physics, Chinese Academy of Sciences, Shanghai 201800, China}

\author{Lie-Wen Chen}
\affiliation{Department of Physics and Astronomy and Shanghai Key
Laboratory for Particle Physics and Cosmology, Shanghai Jiao Tong
University, Shanghai 200240, China}


\author{Dan Cozma}
\affiliation{IFIN-HH, 077125 M\v{a}gurele-Bucharest,
Romania}

\author{Zhao-Qing Feng}
\affiliation{Institute of Modern Physics, Chinese Academy of
Sciences, Lanzhou 730000, China}

\author{Subal Das Gupta
}
\affiliation{Physics Department, McGill University, Montreal,
H3A 2T8, Canada}


\author{Natsumi Ikeno}
\affiliation{Department of Life and Environmental Agricultural Sciences, Tottori University, Tottori 680-8551, Japan}

\author{Che-Ming Ko}
\affiliation{Cyclotron Institute and Department of Physics and
Astronomy, Texas A$\&$M University, College Station, Texas 77843,
USA}


\author{Bao-An Li}
\affiliation{Department of Physics and Astronomy, Texas A$\&$M
University-Commerce, Commerce, Texas 75429-3011, USA}

\author{Qing-Feng Li}
\affiliation{School of Science, Huzhou University, Huzhou 313000,
China}
\affiliation{Institute of Modern Physics, Chinese Academy of
Sciences, Lanzhou 730000, China}

\author{Zhu-Xia Li}
\affiliation{China Institute of Atomic Energy, Beijing 102413, China}

\author{Swagata Mallik}
\affiliation{Physics Group, Variable Energy Cyclotron Centre,
1/AF Bidhan Nagar, Kolkata 700064, India}

\author{Yasushi Nara}
\affiliation{Akita International University, Akita 010-1292, Japan}

\author{Tatsuhiko Ogawa}
\affiliation{Research Group for Radiation Transport Analysis, Japan Atomic Energy Agency, Shirakata, Tokai, Ibaraki 319-1195, Japan}

\author{Akira Ohnishi}
\affiliation{Yukawa Institute for Theoretical Physics, Kyoto University, Kyoto 606-8502, Japan}

\author{Dmytro Oliinychenko}
\affiliation{Frankfurt Institute for Advanced Studies, Johann Wolfgang Goethe University, Ruth-Moufang-Strasse 1, 60438 Frankfurt am Main, Germany}

\author{Massimo Papa}
\affiliation{INFN-LNS, Laboratori Nazionali del Sud, 95123 Catania,
Italy}

\author{Hannah Petersen}
\affiliation{Frankfurt Institute for Advanced Studies, Johann Wolfgang Goethe University, Ruth-Moufang-Strasse 1, 60438 Frankfurt am Main, Germany}
\affiliation{Institute for Theoretical Physics, Goethe University, Max-von-Laue-Strasse 1, 60438 Frankfurt am Main, Germany}
\affiliation{GSI Helmholtzzentrum f\"{u}r Schwerionenforschung, Planckstr. 1, 64291 Darmstadt, Germany}

\author{Jun Su}
\affiliation{Sino-French Institute of Nuclear Engineering $\&$
Technology, Sun Yat-sen University, Zhuhai 519082, China}

\author{Taesoo Song}
\affiliation{Frankfurt Institute for Advanced Studies, Johann Wolfgang Goethe University,
Ruth-Moufang-Strasse 1, 60438 Frankfurt am Main, Germany}
\affiliation{Institut f\"{u}r Theoretische Physik, Universit\"{a}t Gie\ss en, Heinrich-Buff-Ring, 35392 Gie\ss en, Germany}

\author{Janus Weil}
\affiliation{Frankfurt Institute for Advanced Studies, Johann Wolfgang Goethe University, Ruth-Moufang-Strasse 1, 60438 Frankfurt am Main, Germany}

\author{Ning Wang}
\affiliation{Department of Physics and Technology, Guangxi Normal
University, Guilin 541004, China}

\author{Feng-Shou Zhang}
\affiliation{Key Laboratory of Beam Technology and Material
Modification of Ministry of Education, College of Nuclear Science
and Technology, Beijing Normal University, Beijing 100875, China}
\affiliation{Beijing Radiation Center, 100875 Beijing, China}


\author{Zhen Zhang}
\affiliation{Cyclotron Institute and Department of Physics and
Astronomy, Texas A$\&$M University, College Station, Texas 77843,
USA}

\begin{abstract}
Simulations by transport codes are indispensable to extract valuable physics information from heavy-ion collisions. In order to understand the origins of discrepancies between different widely used transport codes, we compare 15 such codes under controlled conditions of a system confined to a box with periodic boundary, initialized with Fermi-Dirac distributions at saturation density and temperatures of either 0 or 5~MeV.
In such calculations, one is able to check separately the different ingredients of a transport code. In this second publication of the code evaluation project, we only consider the two-body collision term, i.e. we perform cascade calculations.
When the Pauli blocking is artificially suppressed, the collision rates are found to be consistent for most codes (to within $1\%$ or better) with analytical results, or completely controlled results of a basic cascade code.
In order to reach that goal, it was necessary to eliminate correlations within the same pair of colliding particles that can be present depending on the adopted collision prescription.
In calculations with active Pauli blocking, the blocking probability was found to deviate from the expected reference values. The reason is found in substantial phase-space fluctuations and smearing tied to numerical algorithms and model assumptions in the representation of phase space.  This results in the reduction of the blocking probability in most transport codes, so that the simulated system gradually evolves away from the Fermi-Dirac towards a Boltzmann distribution. Since the numerical fluctuations are weaker in the Boltzmann-Uehling-Uhlenbeck codes, the Fermi-Dirac statistics is maintained there for a longer time than in the quantum molecular dynamics codes.
As a result of this investigation, we are able to make judgements about the most effective strategies in transport simulations for determining the collision probabilities and the Pauli blocking. Investigation in a similar vein of other ingredients in transport calculations, like the mean field propagation or the production of nucleon resonances and mesons, will be discussed in the future publications.
\end{abstract}


\maketitle

\section{Introduction}
The investigation of properties of nuclear matter away from saturation density, such as the nuclear equation of state (EoS), is a great challenge for nuclear physics. It is also critically important for the understanding of astrophysical objects and processes in neutron stars and core collapse supernovas for a broad range of densities.
Heavy-ion collisions provide a unique opportunity to study the nuclear equation of state in the laboratory, for a range of densities, temperatures, and neutron-proton asymmetries. However, heavy-ion collisions create transient states out of equilibrium, and theoretical methods are needed to infer equilibrium information. The inference in collisions at incident energies between the Fermi-energy regime and several GeV per nucleon normally relies on transport theory. Because of the complexity of transport equations, and in particular their dimensionality, the solution is not sought directly but rather through algorithms that, in particular, invoke statistical sampling and finite phase-space resolutions.

Ideally, the determination of physics quantities from heavy ion experiments should be independent of the utilized implementation of transport theory and details of modeling in the transport code. However, recently it became apparent that different conclusions could be drawn from the same data, with no obvious physics reasons, while relying on transport simulations, e.g.\ in the investigations of isospin equilibration in peripheral collision (isospin diffusion) \cite{Tsang04,Galichet09, Tsang09, Rizzo08}, or in the interpretation of ratios of charged pions \cite{Xiao09, Cozma17, Hong14, Xie13, ZQFeng10, Song15}. On one hand, transport simulations differ in various technical assumptions; on the other hand, the inputs to these simulations are often different in subtle ways even when major physics assumptions are the same. The impacts of these on predictions and conclusions are often difficult to disentangle.  This situation led to the idea of a systematic comparison and evaluation of transport codes under controlled conditions to eventually provide benchmark calculations for transport codes and thus to improve the ability to reach firm conclusions.

This project was started some time ago with dedicated workshops at European center for theoretical studies in nuclear physics an related areas (ECT*) in Trento, Italy \cite{Kolomeitsev05}. It continued through dedicated workshops at the Shanghai Jiao Tong University, China \cite{Xu2016}, and at Facility for Rare Isotope Beams/National Superconducting Cyclotron Laboratory (FRIB/NSCL) in East Lansing, USA \cite{ICNTwebsite}, in addition to various smaller satellite workshops tied to larger meetings. The final result of the Shanghai workshop was a comparison of 19 widely used codes in the field, in simulations of Au + Au collisions at incident energies of 100 and 400 AMeV, for specified input mean field and cross sections, as well as specified impact parameters and initialization details. The compared aspects included the stability of the initialized nuclei, the distributions of the collision probabilities in time and energy, the efficiency of Pauli blocking for the final states, and  predicted flow observables.  The outcome of these investigations was published in 2016 \cite{Xu2016}.
An attempt was made to quantify the model uncertainty in the transverse flow predictions, which was found to depend on the incident energy and amounted to about 30\% at 100 and 13\% at 400 AMeV.  There were indications that a large part of the observed differences in the outcomes resulted from differences in the initialization of the systems and in the treatment of Pauli blocking. Integration of the mean-field equations of motion also seemed to play a role.
However, the origins of the differences are often difficult to pin down unambiguously, since various effects interplay and propagate, and are thus difficult to disentangle. As an example, if the Pauli blocking is less efficient then the reaction becomes more  violent, lower densities are achieved, and the mean field is impacted.

To make progress in understanding the differences among different transport codes observed in Ref.\cite{Xu2016}, we
perform
box calculations, i.e.~simulations of nuclear matter enclosed in a box with imposed periodic boundary conditions. The box calculations have several advantages: (1) the initial conditions are straightforward to realize, (2) the average density remains constant in time, (3) the different aspects of heavy-ion collisions can be isolated and tested separately, e.g.\ the collision probabilities and blocking in cascade calculations
without mean field as reported in this paper,
and the mean field propagation in Vlasov calculations without collision as planned in next installment, and (4) there are in some cases exact limits available from kinetic theory or Landau theory, against which the performance of the codes can be judged rather than against each other.
It is often standard practice also in other fields to subject a code to tests under controlled conditions, including simulations in a box, and to incorporate this as an option in the code,
e.g., to verify that the calculations are consistent with the assumed EoS of symmetric and asymmetric nuclear matter;   see Refs.~\cite{Xu2005,Lepers2010} for some examples outside of intermediate-energy nuclear physics.
However, this is the first time when the performance of different transport codes is compared in box calculations and when consistency with adopted algorithms or assumptions in such a situation is tested independently.

As a word of caution in comparing the different codes against each other and against any known limits, one should keep in mind that: (1) fundamental differences may be present between an idealized kinetic equation and a simulation, such that the idealized limit cannot be reached as a matter of principle; and (2) along that line, the different approaches to transport, namely Boltzmann-Vlasov and molecular dynamics codes, which are briefly described below, start from different philosophies in modeling heavy-ion collisions, and thus one cannot expect that they completely agree with each other. However, differences between codes of the same type and differences with the exact limits in many cases can suggest improvements of the codes. In fact, in the course of the comparisons presented in this paper, some improvements already have been implemented in some of the codes. In other cases, some codes that showed big discrepancies were taken out of the present comparison for further work on them.
Thus, it should be noted that the codes used in this paper may not in all cases and aspects be identical to the versions of the codes compared in Ref.\cite{Xu2016}.

In the present paper, we discuss only the collision term, i.e., we perform cascade calculations without a mean field. In a subsequent paper, we plan to study the mean field evolution, i.e., perform Vlasov calculations without a collision term.
Further possibilities will be discussed in the outlook section.

The treatment of the collision term is the most critical part in solving a transport equation, since physically it determines the dissipation in the system and numerically it accounts for the biggest part of the expense in a calculation. Because of its non-linearity, the collision term cannot be accounted for in any direct solution to the transport equation but is rather integrated stochastically. This is one reason why most of the hidden prescriptions in different implementations of transport programs are made here. The evaluation of the collision term has two main steps, first determining the essentially classical probability that two (test) particles collide and second checking the main quantum ingredient in a transport simulation to determine whether the final states of a collision are allowed by the Pauli principle.  Both of these steps will be discussed in detail below.

We employed the same principal assessment procedures as developed in the context of Ref.\cite{Xu2016}. Contributors of the participating codes performed specified ``homework" calculations. The resulting files were sent to a subgroup of the organizing committee for evaluation. The results were then discussed in a brief meeting after the Sixth International Symposium on Nuclear Symmetry Energy (NuSYM16) in Beijing~\cite{NuSym16}, and thereafter extensively at the Transport 2017 Workshop, embedded in the International Collaboration in Nuclear Theory (ICNT) Program supported by FRIB at Michigan State University (MSU)~\cite{ICNTwebsite}. In between these meetings  updates of the homework were requested from the contributors.

This paper is organized as follow: After a short introduction, a short description of the two families of transport approaches is given in Sec.~\ref{transport approaches}, stating the main differences between the approaches and clarifying the terminology.  The homework specifications pertaining to this paper are described in Sec.~\ref{homework description}.  The results are described in Sec.~\ref{results without blocking} for the collision probabilities and in Sec.~\ref{results with blocking} for the Pauli blocking. Finally, we discuss the results in Sec.~\ref{discussion} and give a summary and an outlook for future work in Sec.~\ref{summary}.

The comparison in Ref.\cite{Xu2016} is the departure point for the present comparison. In this context, it is of interest to gain more insight into the codes than provided by the tables in Ref.\cite{Xu2016}.
Therefore, as a supplement to Ref.\cite{Xu2016} and the present paper, we will make available a compact description of all the codes, which will be submitted for publication as a review paper in the very near future.
All together there are 10 Boltzmann-Vlasov and 11 molecular dynamics type codes. Participating codes in this paper are listed in Table \ref{tab:codes}. Some codes of Ref.\cite{Xu2016} have dropped out while four codes have been added. The antisymmetrized molecular dynamics (AMD) code \cite{AMD} is not included in the present comparison, since a box condition in this code is more involved and not comparable to the treatment in the semi-classical codes.

\begin{table*}[htbp]
\caption{\label{tab:codes} The acronyms, code correspondents, kinematic character (relativistic/nonrelativistic), and representative references of the seven BUU-type and eight QMD-type codes
participating in the present comparison.}
    \begin{tabular}{ccccc}
    \hline
    \hline
    Type & Acronym & Code Correspondents & Rel/Non-Rel & Reference \\
    \hline
  &    BUU-VM\footnote{BUU code developted jointly at VECC and McGill.}& S.~Mallik
     & rel &  \cite{VECC}\\
   &  GiBUU & J.~Weil & rel &  \cite{Gaitanos} \\
   &  IBUU   &  J.~Xu, L.~W.~Chen, B.~A.~Li & rel & \cite{BALi08} \\
BUU   &  pBUU   &  P.~Danielewicz & rel & \cite{Pawel}\\
   &  RVUU   & T.~Song, Z.~Zhang, C.~M.~Ko & rel & \cite{Song15,KoCM}\\
   &  SMASH & D.~Oliinychenko, H. ~Petersen & rel & \cite{Weil}\\
   &  SMF    & M.~Colonna & non-rel & \cite{Colonna} \\
     \hline
 &     CoMD & M.~Papa & non-rel & \cite{Massimo,Pap05}\\
   &  ImQMD\footnote{ImQMD-CIAE in Ref.\cite{Xu2016}.} & Y.~X.~Zhang, Z.~X.~Li & rel &\cite{YXZhang}\\
   &  IQMD-BNU &J.~Su, F.~S.~Zhang & rel & \cite{JunS}\\
   &  IQMD-IMP\footnote{Also known as LQMD in literature.} & Z.~Q.~Feng & rel &\cite{ZQFeng}\\
QMD   
   &  JAM & A.~Ono, N.~Ikeno, Y.~Nara & rel & \cite{Ono}\\
   &  JQMD & T.~Ogawa & rel & \cite{Ogawa}\\
   &  TuQMD & D.~Cozma & rel & \cite{Cozma}\\
   &  UrQMD & Y.~J.~Wang, Q.~F.~Li & rel &\cite{QFLi,Bass} \\
    \hline
    \hline
    \end{tabular}
\end{table*}

\section{Transport approaches}
\label{transport approaches}
In this section we briefly introduce the transport approaches. The purpose here is mainly to establish the concepts and terminology for the discussion that will follow. More detailed remarks can be found in Ref.\cite{Xu2016}.

Transport approaches for heavy-ion collisions can be roughly divided into two families, those of the Boltzmann-Vlasov type, here collectively referred to as Boltzmann-Uehling-Uhlenbeck (BUU) approaches, and those of the molecular dynamics type, here called quantum molecular dynamics (QMD) approaches, consistent with the terminology employed in the literature for their most widely used representatives.

In BUU approaches, the goal is to describe the evolution of the one-body
phase space occupation probability
$f(\vec{r},\vec{p}; t)$  as a function of time under the action of a mean field potential $U[f]$, usually derived from a density functional, and two-body collisions specified by an in-medium cross section $d\sigma^\text{med}/d\Omega$. The
non-relativistic BUU equation reads
\begin{equation}
\Big(\frac{\partial}{\partial t}+ \frac{\vec{p}}{m}\cdot \vec{\nabla}_r-\vec{\nabla}_r U\cdot \vec{\nabla}_p\Big) f(\vec{r},\vec{p};t)=I_{\text{coll}} (\vec{r},\vec{p};t) \, ,
\label{eq:BUU}
\end{equation}
with the collision term
\begin{multline}\label{eq:Icoll}
I_{\text{coll}}=\frac{g}{(2\pi\hbar)^{3}}
 \int d^3p_1d\Omega\, v_\text{rel} \frac{d\sigma^\text{med}}{d\Omega}
\\\times\bigl[f'f_1'(1-f)(1-f_1)- ff_1(1-f')(1-f_1')\bigr],
\end{multline}
where $g$ is the degeneracy,
$f_1$ refers to $f(\vec{r},\vec{p}_1; t)$, primed quantities refer to a relative state at solid angle $\Omega$, and $v_\text{rel}=|\vec{v}-\vec{v}_1|$ is the relative velocity.
The two-body collision, included above in the collision integral, produces a change in momenta, $\vec{p}+ \vec{p}_1 \rightarrow \vec{p}'+\vec{p}_1'$ or reversely, with the phase-space factors $f$
[$f'$] accounting for the
occupation probabilities of initial states in the loss [gain] term.
The Pauli blocking factors $(1-f)$ [$(1-f')$]
describe the statistical ability to populate the final states in the fermionic system in the gain [loss] term. In the case of cascade calculations, the potential term on the left-hand side is absent.
The BUU theory has also been formulated in a relativistic framework, and actually most codes in this comparison use a relativistic formulation. For simplicity and since the potentials are not relevant here, we show the non-relativistic form
with a momentum-independent mean-field potential.
Comments on the relativistic treatment of the collision term are given below. The integro-differential non-linear BUU equation is solved numerically. To this end, the distribution function is represented in terms of finite elements, so-called test particles (TP)~\cite{Wong82}, as
\begin{equation}
f(\vec{r},\vec{p};t)=\frac{(2\pi\hbar)^3}{g N_{\text{TP}}} \sum_{i=1}^{AN_\text{TP}} G(\vec{r}-\vec{r}_i(t)) \, \tilde{G}(\vec{p}-\vec{p}_i(t)) \, ,
\label{eq:fTP}
\end{equation}
where $N_{\text{TP}}$ is the number of test particles (TP) per nucleon
(set to 100 in this work),
$\vec{r}_i$ and $\vec{p}_i$  are the time-dependent coordinates and momenta of the test particles, and $G$ and $\tilde{G}$ are the shape functions in coordinate and momentum space, respectively, with a unit norm (e.g.\ $\delta$ functions or normalized Gaussians).  The degeneracy factor $g=4$ is to define $f(\vec{r},\vec{p},t)$ as the spin-isospin averaged phase space occupation probability.  It is also possible to express the distribution function for each isospin (or spin) state in a similar way.  Upon inserting the ansatz~\eqref{eq:fTP} into the left-hand side of Eq.~\eqref{eq:BUU}, i.e., without the collision integral, Hamiltonian equations of motion for the test particle propagation follow:
\begin{equation}
\frac{d\vec{r}_i}{dt}=\vec{\nabla}_{{p}_i} H \hspace*{2em} \text{and} \hspace*{2em} \frac{d\vec{p}_i}{dt}=-\vec{\nabla}_{r_i} H \, .
\end{equation}
The collision term is accounted for by a Monte-Carlo procedure, that usually allows us to interpret its integration in terms of stochastic collisions between the test particles, according to their cross section and relative distance from each other.
This is explained in more detail in Sec.~\ref{sec:colldesc}. The calculation of the collision term is numerically the most expensive, as it nominally scales with the number of test particles as $(AN_{\text{TP}})^2$. In the full-ensemble method, collisions between all test particle pairs are considered and the cross section is divided by $N_{\text{TP}}$. In the nominally less expensive parallel-ensemble method, the test particles are divided into $N_{\text{TP}}$ ensembles of $A$ test particles each, and collisions are only considered within each ensemble with the full cross section. For calculating the mean field and the Pauli blocking factors, the phase-space distributions are averaged over all ensembles.

In the QMD approach, the evolution of a heavy-ion collision is formulated in terms of the changes in nucleon coordinates and momenta, similar to classical molecular dynamics, but with particles
described by wave packets of finite width.
They move under the influence of nucleon-nucleon interactions, which are usually consistently accounted for by density functionals. The method can also be viewed as derived from the time-dependent Hartree method with a product trial wave function of single-particle states in Gaussian form
\begin{align}\label{eq:QMDwf}
&\Psi(\vec{r}_1,\dots, \vec{r}_A; t) =  \prod_{i=1}^A \phi_i(\vec{r}_i;t), \\
&\phi_i(\vec{r}_i;t) = \frac{1}{[2\pi (\Delta x)^2\big]^{\frac{3}{4}}}\exp\bigg[-\frac{[\vec{r}_i-\vec{R}_i(t)]^2}{4(\Delta x)^2}\biggr]e^{(i/\hbar)\vec{P}_i(t)\cdot\vec{r}_i}. \nonumber
\end{align}
The centroid positions $\vec{R}_i(t)$ and momenta $\vec{P}_i (t)$ are treated as variational parameters within the variational principle for the time-dependent Hartree equation.
The widths $\Delta x$ are kept fixed and thus are not variational parameters, in order for the wave function to be able to describe finite distance structures, as observed in the fragmentation of colliding nuclei.
This strategy yields equations of motion of the same form as in BUU for the coordinates of the wave packets. This method has been extended to include anti-symmetrization in the wave function in the AMD method \cite{AMD}, which makes the equations of motion more complicated but, in principle, similar. In QMD, a stochastic two-body collision term is also introduced and treated in very much the same way as in BUU, but now for nucleons and the full cross section, and Pauli-blocking factors are calculated event by event. Thus the strategies to simulate the collision term, as discussed below, are very similar within the two approaches.

The main difference between the two methods lies in the amount of fluctuations and correlations in the representation of the phase space distribution. In the BUU approach the phase space distribution function is seen as a one-body quantity and a smooth function of coordinates and momenta and it can be approximated better by increasing the number of test particles in the solution. In the limit of $N_{\text{TP}}\rightarrow \infty$, the BUU equation is solved exactly. In this limit the solution is deterministic and does not contain fluctuations.  If fluctuations are considered to be important, as is the case of studying the cluster and fragment production, these may be introduced in a complementary manner, in particular through the Boltzmann-Langevin equation, which adds a fluctuation term on the right-hand side of Eq.~\eqref{eq:BUU}.  This equation is solved approximately in the codes SMF and BLOB \cite{Colonna, Napoli}. Of course, numerical fluctuations are present in practical calculations with a finite number of test particles.

In QMD, the fluctuations are present due to the representation in terms of a finite number of wave packets.  In addition, classical correlations are present, if explicit two-body interactions are used.
Thus, in the philosophy of QMD one wants  to go beyond the mean field approach and include correlations and fluctuations from the beginning. As we will see in these comparisons, this is at the expense of destroying the fermionic character of the system more rapidly and of reverting to a classical system.
The fluctuations in QMD-type codes are regulated and smoothed by choosing the parameter $\Delta x$, the width of the wave packet, cf. Eq.~\eqref{eq:QMDwf}.  Also the collision term, which relocates nucleon wave packets in momentum space, introduces more fluctuations than those for the collision term in BUU.
QMD can be seen as an event generator, where the time evolution of different events is solved independently and therefore the fluctuations among events are not suppressed even in the limit of infinite number of events. The effects of this difference in the amount of fluctuations between the two approaches will clearly be seen in the comparisons that will follow.

Thus fluctuations in transport codes reside partly in the non-smoothness of the phase space distribution and are partly generated by the discrete jumps in momentum space in the simulation of the collision term. The first originate from the representation of the phase space distribution by finite elements. They may be called initial state fluctuations, since they exist in the initialized state and are propagated in the evolution of the collision. However, they are of purely statistical origin and are not determined by physical arguments. These fluctuations are a property of each individual BUU run or QMD event, and are larger, as argued above, for QMD approaches. In addition, there are event-by-event fluctuations in QMD, or fluctuations between different runs in BUU.

\section{Homework description}
\label{homework description}

The box calculations are performed with periodic boundary conditions. Reflecting boundary conditions are not used because they could give rise to edge effects, negligible only in the limit of very large boxes.
In contrast, with periodic boundary conditions the box can be kept relatively small with no significant finite-size effects. The dimensions of the cubic box are $L_\alpha = 20 \, \text{fm}$, $\alpha \equiv x,y,z$. The position of the center of box is ($L_x$/2, $L_y$/2, $L_z$/2).
In a periodic box a particle that leaves the box on one side should enter it from the opposite side with the same momentum.
Once a coordinate $\alpha$ ventures outside of the box, it may be reset with $r_\alpha \rightarrow \mathop{\text{modulo}}( r_\alpha,  L_\alpha )$. Similarly, the separation between two points $\Delta r_{ij,\alpha}=r_{i,\alpha}-r_{j,\alpha}$ must be redefined as $\Delta r_{ij,\alpha} \rightarrow \mathop{\text{modulo}}( \Delta r_{ij,\alpha}+L_\alpha/2, L_\alpha ) - L_\alpha/2$.
This method is completely sufficient and will cope with all structures, as long as the characteristic lengths are shorter than L/2. These lengths are the widths of the wave packets or test particles, and the collision distance $\sqrt{\sigma / \pi}$, which is explained below in Sec. IV.B and represents the interaction range.
This periodic box condition applies only to classical or semiclassical approaches.
In quantum mechanical approaches such as in AMD \cite{AMD}, the implementation of a periodic box calculation is more involved, since now the wave functions have to satisfy the boundary condition, implying that the momenta get discretized in steps of the order of $\delta p= 2\pi/L\approx$ 62 MeV/c, which is not so much smaller than the Fermi momentum. A special code would have to be written for this, which would not be comparable to the semi-classical codes, and would also be very different from the code used for heavy-ion collisions. However, in this box comparison we want to change the codes as little as possible from those used for heavy-ion collisions.

In the box calculations presented in this paper,
we require no nuclear mean field and no Coulomb interactions. We further assume an isotropic constant elastic cross section of $40 \, \text{mb}$. All inelastic processes are turned off. There are no restrictions on the collision times and energies; in particular a particle that has already collided in a time step can collide again with another particle. There is also no threshold for the collision energy.  A time step of $\Delta t = 0.5$ or $1.0 \, \text{fm/}c$ is recommended in integrating the integro-differential equations.

The box is initialized with a uniform density $\rho=0.16 \, \text{fm}^{-3}$, with isospin asymmetry zero. This corresponds to $A=1280$ nucleons, 640 neutrons and 640 protons in the box. Particle positions are initialized randomly from $0$ to $L_\alpha$. In momentum space, we consider two cases, corresponding to matter at rest at the temperature of $T=0$ and $T=5 \, \text{MeV}$. For $T=0$ MeV, the particle momenta are initialized randomly in a sphere with the Fermi momentum and for $T=5$ MeV with the Fermi distribution, $f=1/\big(1+\exp{\big[(\epsilon-\mu)/T)\big]\big)}$, with $\epsilon=p^2/2m$ and relativistically $\epsilon=\sqrt{m^2+p^2}-m$, with the nucleon mass $m=938$ MeV/$c^2$. The chemical potential $\mu$ is obtained from the normalizing condition $\frac{2}{(2\pi\hbar)^3}\int f \text{d}^3 p=\rho_{n,p}$ for neutrons and protons, respectively. In Table ~\ref{tab:mut}, we list the values of $\mu$, the average energy per particle $\overline{\epsilon}$, and the temperature $T_B$ to which the system equilibrates in the Boltzmann limit while conserving energy, for both non-relativistic and relativistic cases at $T=0$ and $5 \, \text{MeV}$. Unfortunately, these precise values of the chemical potentials and the Fermi momentum $p_F=263.04$ MeV/$c$ were not correctly specified in the homework description and it turned out that many codes used slightly different values of these parameters.  On their own, these initialization differences can generate differences in the studied forthcoming collision numbers of up to 2\%, which will be compared in detail in the appendix

\begin{table}[htbp]
\caption{\label{tab:mut}
Values of the chemical potential $\mu$, the average energy $\overline{\epsilon}$, and the temperature $T_B$ for equilibration in the Boltzmann limit for the starting temperatures in the Fermi distribution of $T=0$ and 5 MeV in the relativistic and non-relativistic cases. The values of $\rho=0.16$ $\text{fm}^{-3}$ and $m = 938$ MeV were assumed.  All results are in MeV.\\[-1ex]
}
\begin{tabular}{l ccc ccc}
\hline
\hline
& \multicolumn{3}{c}{$T=0$ MeV} & \multicolumn{3}{c}{$T=5$ MeV}\\
& $\mu$ & $\overline{\epsilon}$ & $T_B$ &
  $\mu$ & $\overline{\epsilon}$ & $T_B$ \\
\hline
Non-relativistic & 36.882 & 22.129 & 14.753 & 36.306 & 23.740 & 15.833\\
Relativistic & 36.184 & 21.827 & 14.284 & 35.544 & 23.510 & 15.364 \\
\hline
\hline
\end{tabular}
\end{table}

In order to gain insights into the consequences of the Pauli-blocking algorithms in the different codes, the simulations are performed with two options.  In option OP1, the default treatment is used, that is the standard for the specific code in simulating heavy-ion collisions.  In
OP2, the Pauli blocking factor, $P_{\text{block}}=(1-f_i )(1-f_j)$, is used with the distribution function $f$ (for the final states $i$ and $j$) calculated from the Fermi-Dirac distribution at the temperature specified for the calculation. Specifically, in the $T = 0$ case the distribution function is equal to 1 or 0 depending on whether the momentum is inside or outside of the Fermi sphere and the number of allowed collisions should be rigorously equal to zero.

The authors of BUU-like codes were asked to provide results from 10 runs with 100 test particles per nucleon and those of QMD codes results from 200 runs.  Two calculational modes are considered, C and CB.  The mode~C is a cascade mode, i.e., no mean field, without Pauli blocking.  The mode CB is a cascade mode with blocking.  When no blocking is employed, obviously the different blocking options, OP1 and OP2, do not apply.  Altogether, with the two modes (C, CB), two blocking options (OP1, OP2) and the two temperatures (T0, T5), there were 6 sets of calculations to be done, named CT0, CT5, CBOP1T0, CBOP2T0, CBOP1T5, and CBOP2T5.

We do not include information on running times of the codes in this comparison. It may be true that the different strategies to perform the simulation, in particular to evaluate the collision term, as discussed in Secs. IV and V, have different calculational expenses. However, this is too much dependent on coding and computers, so we do not consider it very meaningful to compare running times. In addition, nowadays these are often not a limiting factor in the simulations, not in a box and also not usually in heavy-ion collisions. If one wants to obtain rough estimates, rather than using a different code, one usually reduces the statistical significance of the simulation, i.e., fewer events in QMD and fewer test particles in BUU.

\section{Results without Pauli blocking}
\label{results without blocking}

The first task in integrating the collision integral is to determine possible collision partners and to find the probability of a collision within a given time step in the absence of blocking. We denote these as the ``attempted'' collisions. The next task, described in Sec. ~\ref{results with blocking}, is to test the Pauli blocking for the final state of the collision. If the state is not blocked, then this is a ``successful'' collision.  In Subsecs. ~\ref{exact limits}, ~\ref{strategies for collision} and ~\ref{repeated collisions}, we discuss exact limits for the collision rates, the procedures to determine the collision probability in the different codes, and the question of correlations between collisions, respectively. In the final subsection ~\ref{results for no-blocking} we compare the results for the different codes in calculations without Pauli blocking on momentum distributions and collision rates to the exact limits.

\subsection{Exact limits of collision rates}
\label{exact limits}
Without mean field, for a uniform and isotropic distribution $f(\vec{r},\vec{p};t)=f(p;t)$, the Boltzmann equation to be solved here is
$\partial f(p;t)/\partial t =I_{\text{coll}}$ with
\begin{equation}\label{eq:Icoll_noblocking}
I_{\text{coll}}=g \int \frac{d^3p_1}{(2\pi\hbar)^3} \, d\Omega \, v_\text{rel}  \, \frac{d\sigma^{\text{med}}}{d\Omega} \,  \big(f' f_1'-ff_1 \big) \, ,
\end{equation}
Here, the change of momenta in the loss term (the second term in the parenthesis) is $\vec{p} + \vec{p}_1 \rightarrow \vec{p}' + \vec{p}_1'$ and reversely in the gain term.
The expected rate of two-body collisions in the volume is
\begin{align}
\label{eq:dNcoll}
\frac{dN_{\text{coll}}}{dt}&= \frac{A}{2\rho} \, g^2 \int \frac{d^3p\,d^3p_1}{(2\pi \hbar)^6} \, v_\text{rel} \, \sigma^\text{med} \, f(p) \, f(p_1)\nonumber\\
& = \frac{1}{2}A \, \rho \, \langle v_\text{rel} \, \sigma^\text{med} \rangle \, .
\end{align}
The quantity $v_\text{rel}$ and the average depend on the treatment of relativity.
Here `non-relativistic' results pertain to the use of $v_\text{rel} = |\vec{p}/m - \vec{p}_1/m|$ together with the non-relativistic Fermi-Dirac or Boltzmann distributions for~$f(p)$ and~$f(p_1)$.  The `quasi-relativistic' results pertain to the use of the relativistic Fermi-Dirac or Boltzmann distributions and $v_\text{rel} = |\vec{p}/E - \vec{p}_1/E_1|$, where $E=m+\epsilon = \sqrt{m^2+p^2}$.  Finally, in the relativistic case, for the Boltzmann distribution, we have the result
\begin{multline}
\label{eq:RelRate}
\frac{dN_{\text{coll}}}{dt}=\frac{1}{2} \, A \, \rho \,
 \frac{1}{4 m^4 \, T_B \, K_2^2(m/T_B)}\\
\times\int_{2m}^{\infty} d\sqrt{s} \,
 s \, (s-4m^2) \, K_1(\sqrt{s}/T_B) \,
\sigma^{\text{med}} \, ,
\end{multline}
where $K_n$ is the $n$th-order modified Bessel function.

\begin{table*}[htbp]
\caption{\label{tab:analcoll}
Limiting values of collision rates in units of $c/\text{fm}$, in absence of Pauli blocking at the starting temperatures of $T=0$ and $5 \, \text{MeV}$  obtained by different methods.  Specifically non-relativistic, quasi-relativistic, and relativistic results are provided for the starting Fermi distributions and equilibrated Boltzmann distributions (see text). The rows labeled `Cascade' represent the reference values obtained in the basic cascade model,
which is explained in Sec. IV.B, where also $\alpha$ and $\Delta t$ are defined. The last column gives results from the basic cascade calculation under exactly the CT0 and CT5 conditions, with the rate averaged over the time period from 60 to $140 \, \text{fm}/c$}.

\begin{tabular}{llcccccc}
\hline\hline
& & \multicolumn{2}{c}{Fermi} & \multicolumn{2}{c}{Boltzmann}& \multicolumn{2}{c}{Cascade (60-140 fm/$c$)}\\
&  &~~~$T=0$~~~ &$T=5$ MeV&~~~$T=0$~~~ &$T=5 $ MeV& ~~~$T=0$~~~ &$T=5$ MeV\\
\hline
Non-relativistic &Num.\ Int.&118.1\hbox to 0em{
\footnote[1]{Within the displayed accuracy the result of numerical integration is fully consistent with the analytic expression.
}} &122.1&115.9\hbox to 0em{
\footnotemark[1]
}
&120.1  &  &    \\
                 &Cascade
                 &118.2 &122.1&115.9&120.1&122.8&127.3\\
\hline
Quasi-relativistic
                 & Num.\ Int.&115.0&118.8&112.3&116.3    &  &       \\
                 & Cascade   &115.0&118.8&112.3&116.3&119.0&123.2\\
\hline
Relativistic
                 &Num.\ Int.&             &          &111.4&115.4     &  &       \\
                 &Cascade ($\delta t=\alpha\Delta t$) & 114.0&117.8&111.4&115.4&118.1&122.3\\
                 & Cascade ($\delta t=\Delta t$)  & 115.2&119.0&112.7&116.7&118.4&122.7\\
\hline\hline
\end{tabular}
\end{table*}

In Table \ref{tab:analcoll}, we give the calculated collision rates in the three cases of non-relativistic, quasi-relativistic, and relativistic treatment for  Fermi-Dirac distributions of temperature $T=0$ and 5 MeV and for the equivalent Boltzmann distributions; cf.~Table \ref{tab:mut}. For a system initialized with the Fermi-Dirac distribution at $T=0$ MeV and constant cross section, the value is given analytically according to Eq.~(\ref{eq:dNcoll}) with $\langle v_\text{rel}\rangle_{\text{Fermi}}=(36/35)(p_{\text{F}}/m)$ and
$\langle v_\text{rel}\rangle_{\text{Boltzmann}}=(4/\sqrt{5\pi})(p_{\text{F}}/m)$. In some other cases, the rate can be obtained by numerical evaluation of the corresponding integrals \cite{thanksJX}.

Some points should be made regarding these limiting values to which the performance of the different codes will be compared below. First, Table \ref{tab:analcoll} shows that the non-relativistic and the two relativistic treatments are different, and thus the codes should be compared to the limits that represent the corresponding intentions in the codes.
Secondly, it is seen that the collision rates for a Fermi-Dirac and a Boltzmann distribution are not very different, with the latter being somewhat lower. For the $T=0$ distributions, the analytical results are given in the last paragraph, which differ by just 2\%. The important point is, that the distributions are normalized to the same total energy. Then the moments of relative velocity $\langle|(\vec{v}_1-\vec{v}_2)|^n\rangle$ have to be identical for $n=0$ (normalization) and $n=2$ (total energy). Then it is also plausible that they are not very different for $n=1$.
This will be of interest later on, since in the evolution of the system the momentum distribution changes from an initial Fermi-Dirac to a Boltzmann distribution.
The values labeled "Cascade" in Table \ref{tab:analcoll} represent reference results obtained from a completely controlled basic transport code,
the details of which are given at the end of Subsec. IV.B.
 It is used here to check against the analytical and numerical results, which are seen to be reproduced essentially exactly.

\subsection{Strategies for collision attempts\label{sec:colldesc}}
\label{strategies for collision}

\begin{table*}[htbp]
\caption{\label{tab:coll}
Characteristics of collision procedures in different codes and a basic cascade code (for a detailed explanation see text).
The notation ``$P=x$'' stands for a condition that is satisfied randomly with the probability $x$.  Comma-separated conditions stand for the logical conjunction.
Quantities with asterisks are in the two-particle center-of-mass frame, while those without asterisk are in the calculational reference frame.
}
\begin{tabular}{lccl}
\hline\hline
 & Distance condition & Time condition & Collision order\\
\hline
\textbf{BUU-type} & & & \\
BUU-VM
& $\pi d_\perp^{*2}<\sigma$
& $|t_{\text{coll}}^*-t_0^*| < \frac12\Delta t$
& fixed order
\\
GiBUU
& $\pi d_\perp^{*2}<\sigma /N_{\text{TP}}$
& $|t_{\text{coll}}^*-t_0^*| < \frac12\Delta t$
& fixed order
\\
IBUU
& $\pi d_\perp^{*2}<\sigma$
& $|t_{\text{coll}}^*-t_0^*| < \frac12\Delta t$
& fixed order
\\
pBUU
& $i,j\in \mbox{the same $V_{\text{cell}}$ volume}$
& $P=\frac{\sigma}{N_{\text{TP}}}\frac{1}{\gamma V_{\text{cell}}}v_{ij}^*\alpha\Delta t$
& randomly nominate $(i,j)$ pairs
\\
RVUU
& $\pi d_\perp^{*2}<\sigma_{\text{max}}/N_{\text{TP}}$, $P=\sigma/\sigma_{\text{max}}$
& $|t_{\text{coll}}^*-t_0^*| < \frac12\Delta t$
& fixed order
\\
SMASH
& $\pi d_\perp^{*2}<\sigma /N_{\text{TP}}$
& $t_{\text{coll}}^{\text{(ref)}} \in [t_0,\ t_0+\Delta t]$
& ordered by $t_{\text{coll}}^{\text{(ref)}}$
\\
SMF
& 
$j=\mbox{closest to $i$ in same ensemble}$
& $P=\frac{1}{2}  \sigma {v}_{ij} \rho_i \Delta t$
& cyclic with random starting for $i$
\\
\hline
\textbf{QMD-type} &  &  & \\

CoMD
& $j = \mbox{closest to $i$}$, $j>i$
\footnote{
In CoMD, the collision is skipped if the particle $i$ or $j$ has already experienced a collision in the same time step.
}
& $P=1-e^{-\sigma {v}_{ij} \rho_i \Delta t}$
& cyclic with random starting for $i$
\\
ImQMD
& $\pi d_\perp^{*2}<\sigma$
& $|t_{\text{coll}}^*-t_0^*| < \frac12\gamma\Delta t$
& fixed order
\\
IQMD-BNU
& $\pi d_\perp^{*2}<\sigma$
& $|t_{\text{coll}}^*-t_0^*| < \frac12  \Delta t$
& fixed order
\\
IQMD-IMP
& $\pi d_\perp^{*2}<\sigma$
& $|t_{\text{coll}}^*-t_0^*| < \frac12 \gamma  \Delta t$
& fixed order
\\
JAM
& $\pi d_\perp^{*2}<\sigma$
& $\bar{t}_{\text{coll}}\in [t_0,\ t_0+\Delta t]$
& ordered by $\bar{t}_{\text{coll}}$
\\
JQMD
& $d_\perp^*<b_{\text{max}}$, $P=\sigma/\pi b_{\text{max}}^2$
& $|\bar{t}_{\text{coll}}-t_0|<\frac{1}{2}\Delta t$
& fixed order
\\
TuQMD
& $\pi d_\perp^{*2}<\sigma$
& $t_{1-}^*, t_{2-}^* < t_{\text{coll}}^* < t_{1+}^*, t_{2+}^*$
& randomly ordered
\\
UrQMD
& $\pi d_\perp^{*2}<\sigma$
& $t_{\text{coll}}^{\text{(ref)}} \in [t_0,\ t_0+\Delta t]$
& ordered by $t_{\text{coll}}^{\text{(ref)}}$
\\
\hline
\textbf{Basic cascade} &  &  & \\
rel.($\delta t=\alpha\Delta t$)
&$\pi d_\perp^{*2}<\sigma$
& $|t_{\text{coll}}^*-t_0^*| < \frac12\alpha\Delta t$
& fixed order
\\
rel.($\delta t=\Delta t$)
&$\pi d_\perp^{*2}<\sigma$
& $|t_{\text{coll}}^*-t_0^*| < \frac12\Delta t$
& fixed order
\\
quasi-relativistic
&$\pi d_\perp^{\text{(ref)}2}<\sigma$
&  $|t_{\text{coll}}^{\text{(ref)}}-t_0| < \frac12\Delta t$
& fixed order
\\
\hline\hline
\end{tabular}
\end{table*}

We now discuss how the collision probability is determined in the simulation codes. In most codes, each pair of (test) particles within an ensemble is tested for a collision at every time step.  Let us consider the possible collision between particles 1 and 2, specified with $(t_0,\vec{r}_1)$ and $(E_1,\vec{p}_1)$, and $(t_0,\vec{r}_2)$ and $(E_2,\vec{p}_2)$, respectively, at the current time $t_0$  in the reference frame of the box (which we call the calculational reference frame).  Generally, there are two necessary conditions for the collision to occur during the
time step.  The minimum distance $d_\perp$ should be within the range of the interaction, and it should be realized at a time $t_{\text{coll}}$ during that time step. This procedure is often called the Bertsch prescription, since it was first formulated in the Bertsch-Das Gupta review article \cite{Bertsch88}.
One should note that the positions and momenta of the particles refer to the centroids of the wave packets $R_i(t)$ and $P_i(t)$ in QMD, respectively, and in BUU to the centroids of finite-size test particles, if those are used. This is reasonable, since inclusion of the width of the wave packet would correspond to an unphysical increase of the interaction range, which would be particularly unrealistic in QMD, where the particles represent nucleons.

The characteristics of the collision procedures in the different codes are collected in Table \ref{tab:coll}, which will be explained in this subsection.  The second and third columns show the distance and time conditions, respectively, for the collision to occur in a given time step.  The last column shows how the order of two-particle pairs is established to check for collisions.

There is not much ambiguity in the condition for the distance $d_\perp$. At the point of the closest approach, the separation of particles 1 and 2 is the purely transverse vector $\vec{d}_\perp$ with regard to the relative velocity vector of the particles.
In the center-of-mass frame of the two particles, their trajectories $\vec{\mathcal{R}}_i^*(t^*)$ without a mean field are straight lines pointing along the constant velocities $\vec{v}_1^*=\vec{p}_1^*/E_1^*$ and $\vec{v}_2^*=\vec{p}_2^*/E_2^*$. The asterisks represent quantities in the two-particle center-of-mass frame, while quantities without asterisk are in the calculational reference frame.  As the trajectories of 1 and 2 are known exactly, the minimum distance can be calculated as
\begin{equation}
\label{eq:dperp*}
d_\perp^{*2} = (\vec{r}_1^*-\vec{r}_2^*)^{2}
-\frac{[(\vec{r}_1^*-\vec{r}_2^*)\cdot\vec{v}_{12}^*]^2}{{v}_{12}^{*2}} \, ,
\end{equation}
with $\vec{v}_{12}^*=\vec{v}_1^*-\vec{v}_2^*$.  In most of the QMD-type codes, the distance condition for a collision to occur is $\pi d_\perp^{*2}<\sigma$, while it is $\pi d_\perp^{*2}<\sigma/N_{\text{TP}}$ in BUU-type codes with the full ensemble method, given that there are $N_{\text{TP}}$ times more particles available for collisions (GiBUU, pBUU, RVUU, SMASH). However, it is also possible to set other conditions such as $\pi d_\perp^{*2}<\sigma_{\text{max}}$ together with the probability $P=\sigma/\sigma_{\text{max}}$, where $\sigma_{\text{max}}$ is an arbitrary constant that must be larger than any possible value of the actual cross section $\sigma$. This is the method recommended in the paper by Bertsch et al.~\cite{Bertsch88}.

It is necessary to know when the two particles will collide, in order to judge whether the collision occurs during the time interval of the current time step. There are various choices of the methods which are described in this paragraph for completeness, though we will eventually see that different choices result in a difference in the collision rates of the order of 1\% for the comparison in this paper. In most of the relativistic codes, the time of the closest approach is considered in the two-particle center-of-mass frame, where it may be written as
\begin{equation}
\label{eq:tcoll*}
t_{\text{coll}}^{*}=t_0^*
-\frac{(\vec{r}_1^*-\vec{r}_2^*)\cdot\vec{v}_{12}^*}{{v}_{12}^{*2}} \, ,
\end{equation}
corresponding to the minimum distance $d_\perp^*$ given by Eq.~(\ref{eq:dperp*}).  Note that $\vec{r}_1^*$ and $\vec{r}_2^*$ are the positions at different times $t_1^*$ and $t_2^*$, respectively, in this frame. In Eq.~(\ref{eq:dperp*}), $\vec{r}_1^*$ and $\vec{r}_2^*$ can actually be any spatial points on the free-propagating trajectories $\vec{\mathcal{R}}_1^*(t^*)$ and $\vec{\mathcal{R}}_2^*(t^*)$, respectively, while Eq.~(\ref{eq:tcoll*}) is valid only when $t_0^*$ is chosen by the condition $\vec{\mathcal{R}}_1^*(t_0^*)-\vec{\mathcal{R}}_2^*(t_0^*)=\vec{r}_1^*-\vec{r}_2^*$~\footnote{
In fact, this condition is satisfied for $t_0^*=(E_2^* t_1^*+E_1^* t_2^*)/(E_1^*+E_2^*)$.  For two particles with the same mass, it is $t_0^*=\frac{1}{2}(t_1^*+t_2^*)$.
}.  In the Bertsch prescription \cite{Bertsch88}, the condition of the closest approach for this time step is set as $|(\vec{r}_1^*-\vec{r}_2^*)\cdot\vec{v}_{12}^*/{v}_{12}^{*2}|<\frac{1}{2}\delta t$ which is equivalent to $t_{\text{coll}}^*\in[t_0^*-\frac{1}{2}\delta t,\ t_0^*+\frac{1}{2}\delta t]$.  Many codes choose $\delta t=\Delta t$, i.e., the same as the time step in the calculational reference frame.  However, a more suitable choice of $\delta t$ is found to be $\delta t=\alpha\Delta t$ \footnote{
  When particles are moving with constant velocities and a time step $\Delta t$ has elapsed in the calculational reference frame, the first term ($t_0^*$) and the second term in Eq.~(\ref{eq:tcoll*}) change by the same amount $\alpha\Delta t$ so that $t_{\text{coll}}^*$ does not depend on the time.  One should choose $\delta t=\alpha\Delta t$ in the Bertsch prescription to be sure that the time step condition for the closest approach is satisfied in exactly only one of the time steps.
}
 where $\alpha$ is defined as
\begin{equation}
\alpha=\gamma \frac{E_1^*E_2^*}{E_1E_2}
\end{equation}
with Lorentz factor $\gamma=1/\sqrt{1-\beta^2}$ where $\beta$ is the velocity of the center-of-mass of the colliding pair. We have the usual time dilation factor $\alpha=1/\gamma$ in the limit that the two particles have a common velocity.  A similar relativistic treatment is made in TuQMD, where the segments of trajectories of both particles for $t\in[t_0-\frac{1}{2}\Delta t,\ t_0+\frac{1}{2}\Delta t]$ in the calculational reference frame are Lorentz transformed to those in the two-particle center-of-mass frame, $\vec{\mathcal{R}}_1^*(t^*\in[t_{1-}^*,t_{1+}^*])$ and $\vec{\mathcal{R}}_2^*(t^*\in[t_{2-}^*,t_{2+}^*])$, for which the closest distance condition is considered.  There is yet another class of codes where the collision points $[t_{\text{coll}}^*,\vec{\mathcal{R}}_1^*(t_{\text{coll}}^*)]$ and $[t_{\text{coll}}^*,\vec{\mathcal{R}}_2^*(t_{\text{coll}}^*)]$ are transformed to the calculational reference frame in which the time coordinates $t_{\text{coll},1}$ and $t_{\text{coll},2}$ are different.  For deciding on a collision in a given time step, the average value $\bar{t}_{\text{coll}}=\frac{1}{2}(t_{\text{coll},1}+t_{\text{coll},2})$ is used, e.g.\ $\bar{t}_{\text{coll}}\in[t_0,\ t_0+\Delta t]$ in JAM and JQMD.  In contrast, in UrQMD and SMASH, the time of the closest approach observed in the calculational reference frame
\begin{equation}\label{eq:tcollref}
t_{\text{coll}}^{\text{(ref)}}=t_0
-\frac{(\vec{r}_1-\vec{r}_2)\cdot\vec{v}_{12}}{{v}_{12}^2}
\end{equation}
is used to judge whether the collision occurs in the current time step. Note that closest approach is a frame dependent concept, so $t_{\text{coll}}^{\text{(ref)}}$ is not the Lorentz transformed quantity of $t_{\text{coll}}^*$.

A different approach, with respect to the closest distance method
discussed above, is employed in pBUU, SMF, and CoMD.
In pBUU, the collision integral is averaged over the spatial volume
$\Delta V$ of the cell where a test particle is located and it is
integrated over time step $\Delta t$.  Thus, the collisions result from
Monte-Carlo integration of the collision integral where the
test particles in the cell sample the phase space distribution in the
initial state of a collision.  The collision can occur between any two
particles in a cell no matter whether they move toward each other or
not.  The collision probability within $\Delta t$ between any pair is
usually very low.  To prevent excessive sampling of potential collisions
that are never completed, a subsample of pairs is nominated for
potential collisions and they are tested for collisions at enhanced
probability.
SMF and CoMD essentially follow the same procedure, but only consider,
in the Monte-Carlo integration, one collisional partner $j$ for each
particle $i$, which is chosen as the closest particle to $i$.
This procedure can also be seen as a mean-free-path method~\cite{Bonasera94}. The mean free path of a particle is determined locally as $\lambda=1/\rho(r)\sigma$ and the collision time between the two test particles with relative velocity $v_{ij}$ is $\tau_{\text{coll}}\approx\lambda/v_{ij}$. Then the probability for the two test particles to collide in the time step $\Delta t$ is chosen as proportional to $\Delta t/\tau_{\text{coll}}=\rho\sigma v_{ij} \Delta t$. The exact expressions for the collision probability for these codes are given in Table~\ref{tab:coll}.

In many codes, the collisions within a given time step are processed for the particle pairs in a fixed order following the originally chosen particle indices.  Alternatively, the order may be scrambled at the beginning of every time step.  Each pair is considered only once in a time step in these codes.  By contrast, in some codes (UrQMD, JAM, SMASH), the collisions within a time step are processed according to the order of the established collision times.
In these codes, after every successful collision, the list of possible future collisions is suitably updated. Without a mean field, the time step $\Delta t$ could in principle be quite long, even taking the evolution until the end of a simulated reaction event.

Of course, in a relativistic treatment the ordering of the collision times is frame dependent, and can be different in the calculational frame and in the rest frame of each particle. Thus it can happen that for a particle, which scatters with a particle 2 and subsequently with a particle 3 in the calculational frame, the second collision occurs before the first in the rest frame of particle 1. Such problems were discussed by Kortemeyer et al.,~\cite{Kortemeyer95}; however, for ultra-relativistic  collisions of center-of-mass energies of the order of 100 AGeV. In the box calculation, where typical energies are the Fermi energy of about 35 MeV, and even in heavy-ion collisions around and below 1 AGeV, as discussed in Ref.\cite{Xu2016}, this will happen very rarely. Most relativistic codes and also our basic cascade code do not take this possibility into account, but some codes (e.g., JAM) eliminate a collision in such a situation, which can reduce the collision rate by $\sim 1\%$ in the conditions of the present comparison.

We test variations of collision prescriptions by employing a basic cascade code written specifically for this purpose~\cite{thanksAO}.  The code calculates possible collisions for all the pairs of particles in each time step of $\Delta t=0.5$ fm/$c$ in a similar way to the Bertsch prescription, i.e., following the condition of the closest approach.  The code runs both in relativistic and non-relativistic kinematics without any mean field.  Different versions of relativistic calculations can be done with different choices of $\delta t$ as listed in the bottom part of Table~\ref{tab:coll}.  The quasi-relativistic case, corresponding to the numerical integration in Eq.~(\ref{eq:dNcoll}) with $v_{12}=|\vec{p}_1/E_1-\vec{p}_2/E_2|$, can be simulated with the relativistic kinematics and initialization, but with an unusual collision condition of the closest approach in the calculational reference frame, using the minimum distance
\begin{equation}
d_\perp^{\text{(ref)}2}
 = (\vec{r}_1-\vec{r}_2)^{2}
-\frac{[(\vec{r}_1-\vec{r}_2)\cdot\vec{v}_{12}]^2}{{v}_{12}^{2}},
\end{equation}
and the corresponding time given by Eq.\ (\ref{eq:tcollref}).  This code was used to simulate the numerical integration of Eq.~(\ref{eq:dNcoll}) by only counting the number of attempted collisions without actually scattering the particles and by initializing the particles at every time step with Fermi-Dirac or Boltzmann distributions.  A factor of 1280/1279 was applied to the obtained collision rate in order to correct for the difference in the self-collisions included in Eq.\ (\ref{eq:dNcoll}) but excluded in the cascade calculation. The obtained collision rates are given in the rows `Cascade' in Table \ref{tab:analcoll} for each of the cases for Fermi-Dirac and Boltzmann distributions. They agree essentially exactly with the results of numerical integration in all the cases of non-relativistic, quasi-relativistic and relativistic ($\delta t=\alpha\Delta t$) treatments. Some deviation of about 1\% is seen in the relativistic case when the time dilation effect is ignored in the calculation with $\delta t=\Delta t$.  Thus, the collision prescription
with the closest approach condition
is quite sufficient for a rather accurate reproduction of the collision rate in this simple case without mean field.

\subsection{Repeated collisions and higher-order correlations}
\label{repeated collisions}
To understand the issue of correlations between collisions, the basic cascade code was run for the prescribed simulation with two-particle scatterings in the homework condition of Fermi-Dirac initialization. The averaged collision rates for the interval 60-140 fm/$c$ are shown in the last columns in Table~\ref{tab:analcoll}. It will be seen below that by those times the system has essentially equilibrated to the Boltzmann distribution. It is noticed that the equilibrated collision rates from the code are higher than the reference values for a pure Boltzmann distribution, to which they should be compared. In fact, when this code is applied with the above Bertsch prescription without any restriction on the particle pairs for which the closest approach condition is tested, the equilibrium collision rates are much higher, i.e., around 150-170 $c$/fm, depending on the time step.  Similar results are also found in many of the transport codes compared here. The reason for these very high rates is that the naively applied Bertsch prescription does not preclude that particles that just collided may collide again in the subsequent time steps.  The latter happens if the final velocities point toward each other, which happens 50\% of the time in the chosen statistical procedure.  This, however, should not occur as an independent process. As particles remain within their interaction range, physically they are within the same collision process that has been accounted for in the single abrupt change in the momenta. Or, expressing this in another way, the $T$-matrix, used to describe cross sections, accounts for ladder diagrams.  The Boltzmann equation, for which the collision integral is evaluated, assumes collisions that are independent of each other and are not repeated.  Thus, for a simulation code seeking a solution of the Boltzmann equation, a condition is needed which prohibits repeated collisions of the same pair of particles.  We call this a corrected Bertsch prescription.  The collision rates from the basic cascade code with the corrected prescription are the ones given in the third column in Table \ref{tab:analcoll}.  They are now closer to the reference values listed under the Boltzmann distribution, but, in fact, they do not agree with them exactly.  We believe that higher-order correlations are responsible for the remaining discrepancies.  For example, after a pair of (test) particles has collided, one of them could collide with some other particle and then the same pair could collide again. This also constitutes a correlation between collisions which is not considered in the Boltzmann equation and not accounted for in an exact evaluation of the collision integral.  It would be increasingly complex to eliminate all these higher-order correlations in a simulation code. This discussion shows that it is not trivial to meet the exact limits in simulations, due to existence of residual correlation between collisions.  Of course, when we consider Pauli blocking of collisions in the next section, most of the repeated collisions will be forbidden, but it is desirable to understand this issue and to modify the collision criteria accordingly.

\subsection{Results for no-blocking simulations}
\label{results for no-blocking}
\begin{figure*}[htpb]
\includegraphics[width=0.7\textwidth]{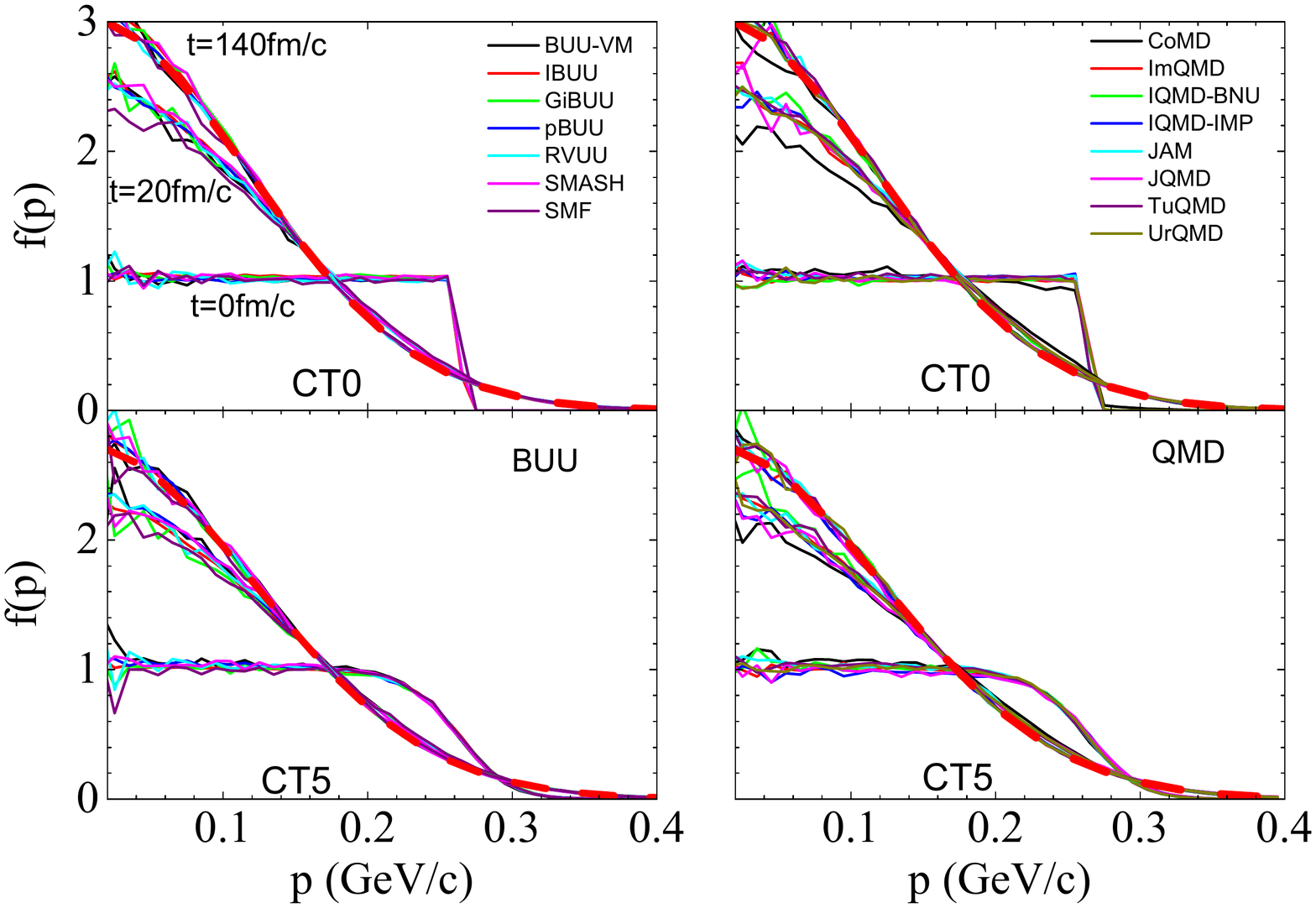}
\caption{(Color online) Momentum distributions at time $t=0$, 20, and $140 \, \text{fm}/c$ for simulations without Pauli blocking initialized with Fermi-Dirac distributions of $T=0$ (upper panels) and $5 \, \text{MeV}$ (lower panels) for the different BUU codes (left panels) and QMD codes (right panels). The thick dashed lines (red) represent ideal non-relativistic Boltzmann distributions for $T_B = 14.753$ and $15.833 \, \text{MeV}$ corresponding to $T=0$ and $T=5\, \text{MeV}$, respectively.} \label{init-np}
\end{figure*}

In this subsection, we discuss the results for collision rates without Pauli blocking for the different simulation codes. The modes CT0 and CT5 are initialized  in terms of Fermi-Dirac distributions at temperatures of 0 and 5~MeV, respectively.  Figure \ref{init-np} shows the momentum distributions at $t=0$, 20, and 140~fm/$c$ for $T=0$ (upper panels) and 5~MeV (bottom panels) from the calculations of BUU-type models on the left and of QMD-type models on the right. The initial distributions agree with the desired one within the expected statistical errors. As expected, the distributions finally reach the respective classical Boltzmann distributions characterized by temperatures given by energy conservation (Table \ref{tab:mut}). The evolution toward equilibrium progresses at a rather rapid pace.  The distributions from different codes cluster together at $t = 20\ \text{fm}/c$ and  all are close to the equilibrated Boltzmann distribution by $t=140$ fm/$c$.

\begin{figure*}[htpb]
\includegraphics[width=0.7\textwidth]{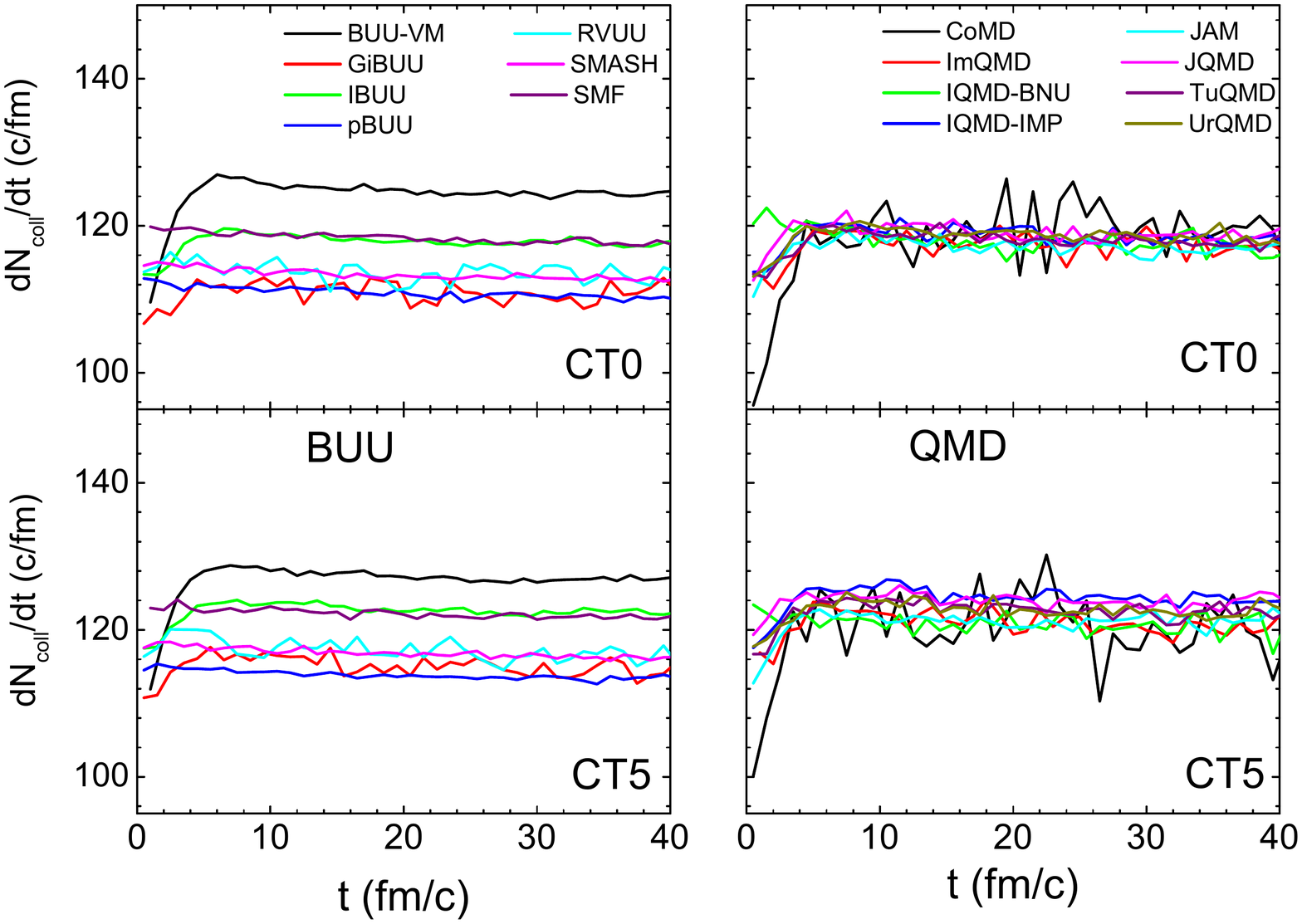}
\caption{(Color online) Time evolution of the collision rate $dN_\text{coll}/dt$ without Pauli blocking for  initializations with $T=0$ (upper panels) and $T=5 \, \text{MeV}$  (lower panels) for BUU codes (left panels) and QMD codes (right panels).  The codes are identified in the legend.
}
 \label{avcoll-t}
\end{figure*}

The time evolution of the collision rates $dN_{\text{coll}}/dt$ in the calculations from the different codes is shown in Fig.~\ref{avcoll-t}. Looking across the codes, the rates are found to settle rather quickly, within about 10~fm/$c$, to equilibrium values, except for CoMD which has large fluctuations. This is consistent with the pace of the changes in the momentum distributions in Fig.~\ref{init-np} from Fermi-Dirac to Boltzmann form.  For most codes, the early collision rates are consistent with Fermi gas expectations in Eq.~\eqref{eq:dNcoll} and Table~\ref{tab:analcoll}.  As time evolves, however, for some codes the rates grow and for some codes the rates change little or slightly drop. By examining Table~\ref{tab:analcoll}, it is observed that, in fact, the rates should drop only very slightly from the initial Fermi-Dirac to the equilibrated Boltzmann distributions. The fact that they rise in some codes, particularly for QMD-type codes, is likely due to higher-order correlations that build up with time.  At the late stage persistent differences between equilibrium rates among different codes or groups of codes are apparent.  Given the offset for the vertical scale in Fig.~\ref{avcoll-t}, they are not large, but still worth illuminating further.

\begin{figure*}[htpb]
\includegraphics[width=0.7\textwidth]{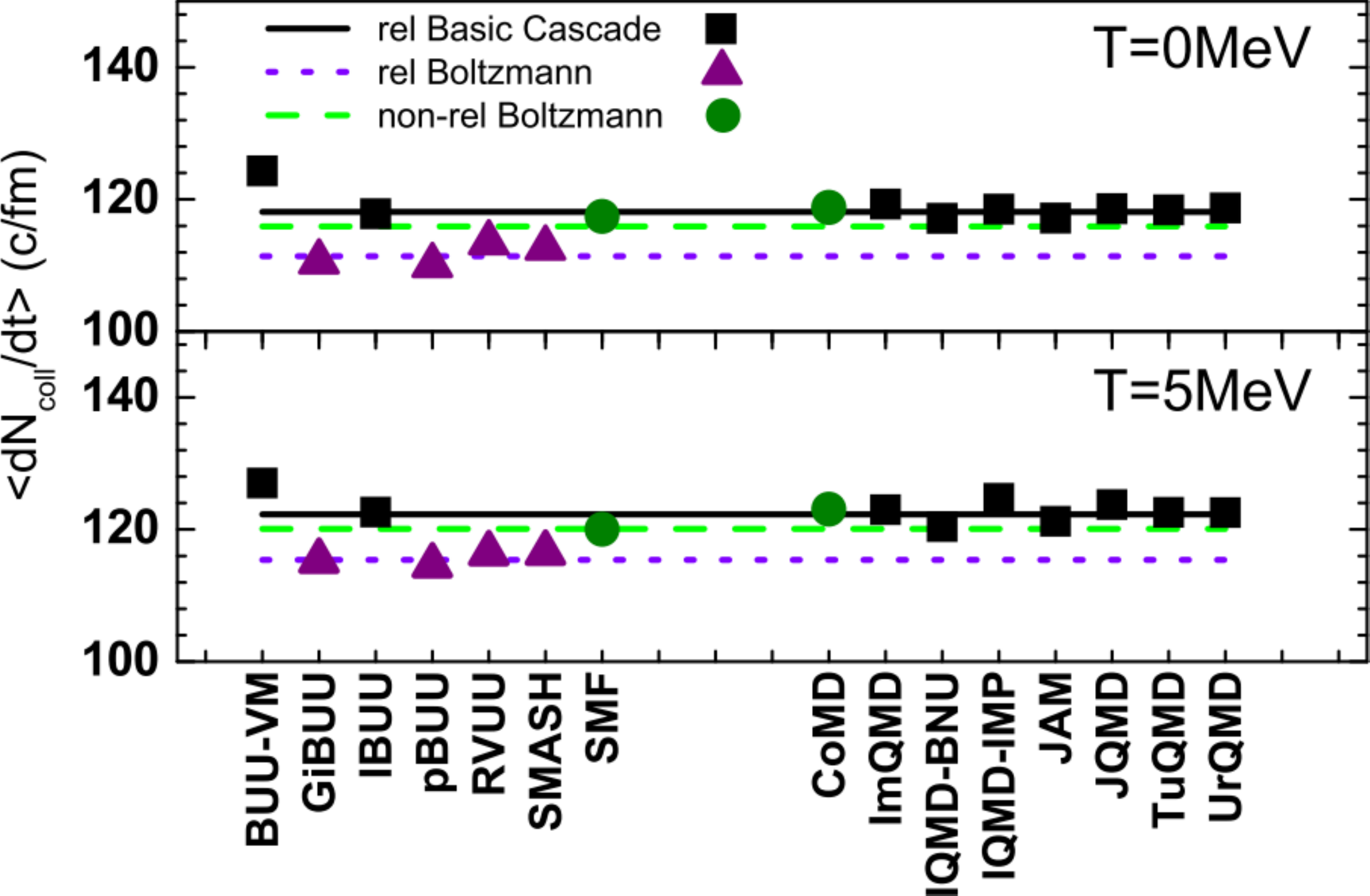}
\caption{(Color online) Collision rates, for systems initialized at $T=0$ (CT0, top, solid symbols) and  $T=5 \, \text{MeV}$ (CT5, bottom, open symbols), for BUU-type codes (left) and QMD-type codes (right).
The lines represent the reference values given in table III: dashed (green) and dotted (purple) the non-relativistic and relativistic Boltzmann rates, respectively; solid (black) the basic relativistic cascade code results. The symbols show the results of the codes averaged over time interval from $60 \, \text{fm}/c$ to $140 \, \text{fm}/c$, cf.\ Fig.~\ref{avcoll-t}. The symbols (and their colors) indicate to which reference values the results of a code should be compared: circle (green) and triangle (purple) to the non-relativistic and relativistic Boltzmann, respectively; square (black) to the relativistic cascade (see text).
} \label{avcoll-num}
\end{figure*}

The equilibrated collision rates averaged over time in the interval  from 60 to $140 \, \text{fm}/c$ for the different transport codes are shown in Fig.~\ref{avcoll-num}, for $T=0$ (upper panels) and $T=5\ \text{MeV}$ (lower panels) initializations.  It is seen that the collision rates agree reasonably well among the different codes, but they do not agree exactly. In the figure, we show as lines the reference values of Table~\ref{tab:analcoll}.
Thus the dashed (green) and dotted (purple)
lines in Fig.~\ref{avcoll-num} represent the values for the non-relativistic and relativistic Boltzmann distributions, respectively. The solid (black) line is the result for the equilibrated rate from the basic cascade code calculation in the relativistic case. The shape (and color) of the symbols indicates, to which reference value the code results should be compared.

Except for SMF, CoMD, and pBUU (see Subsec. IVB), the represented codes use the corrected Bertsch procedure (prior to this box comparison some codes used the original Bertsch prescription and showed correspondingly much higher rates).
The results for the codes should be compared to the reference values corresponding to the treatment of dynamics and of relativity specific for the code, giving hopefully a way to understand the remaining differences.   For example, the rates from BUU codes that rely on the full ensemble method (GiBUU, pBUU, SMASH, and RVUU) tend to coincide with those from the relativistic Boltzmann distribution, around $111\ c/\text{fm}$ in the case of $T=0$ initialization. This can be understood since cross sections are reduced in the full ensemble method and the chances of scattering back into the vicinity  of a preceding collision partner are correspondingly reduced, i.e.\ effects of higher-order correlations are suppressed.  The rates from QMD codes and from the BUU codes that rely on the parallel ensemble method (e.g., IBUU) tend to agree with those from the relativistic version of the basic cascade model, i.e., around $118\ c/\text{fm}$.  In the parallel ensemble method the correlations are thus similar to those in QMD dynamics.  BUU-VM shows unexpectedly high collision rates. As discussed above SMF, CoMD and in a more general sense also pBUU used the mean-free-path method, and thus they come close to the corresponding exact limit, in these cases the non-relativistic Boltzmann value.

As briefly mentioned in Sec. ~\ref{homework description}, there can be slight differences in the initialization between the codes, due to an inprecise specification in the homework. These contribute to the differences seen in Fig. 3. In the appendix, the differences are analyzed more precisely and an attempt is made to separate these from the intrinsic differences of the codes. The appendix describes a more precise analysis of the results of this section. However, it does not affect the global view on the results.

In the global view, the collision rates without blocking in Fig.~\ref{avcoll-num} are rather close to the reference values (within a few percent) and the differences between the codes, resulting from different treatments (relativity, remaining correlations) are of the order of 10\%. Such differences would not essentially affect the result of a simulation of heavy-ion collisions. Thus we can say that the treatment of collision probabilities is well under control. In the next section, we will discuss how this changes when Pauli blocking is included.

\section{Results with Pauli Blocking}
\label{results with blocking}

The Pauli blocking is crucial in a transport calculation. Without it, the distribution very quickly reverts to a Boltzmann distribution. In this section, we compare the box cascade calculations including the Pauli blocking factors in the collision term in Eq.~(\ref{eq:Icoll}). We discuss the option 1 modes CBOP1T0 and CBOP1T5, where the Pauli blocking is handled as in the normal use of the codes, and the option 2 modes CBOP2T0 and CBOP2T5, where the Pauli blocker is always taken as the initial Fermi-Dirac distribution at $T=0$ and 5 MeV, respectively.

\subsection{Phase-space occupation probabilities}
\label{phase space occupation}

To take into account the Pauli blocking, one has to calculate the phase-space occupation probabilities in the final state of a collision of two (test) particles. This then determines the blocking factors $(1-f')(1-f'_1)$ in the loss term of Eq.~(\ref{eq:Icoll}) and similarly for the gain term.  Different prescriptions are used in different codes to do this.
The calculation of the occupation probabilities involves some kind of averaging. This is usually done by spreading particles in addition to a range in position also over some range in momentum, and by determining the contribution to the local phase-space blocking by some overlap of distributions tied to the neighboring test particles.
This can be seen as a sampling of the phase space with finite resolution, i.e. as an effect of coarse-graining. This was studied by Abe et al., \cite{Abe96}, who showed that it  effectively introduces classical dissipation, and thus drives the system to a classical distribution.

In the molecular dynamics codes ImQMD, IQMD-BNU, JAM, JQMD, and UrQMD, the occupation probability $f_i'$ at the centroid of the scattered wave packet with final momentum $P_i'$ is obtained from the Wigner distribution function corresponding to the QMD wave function given in Eq.~(\ref{eq:QMDwf}), with the self-contribution excluded, i.e.,
\begin{align}\label{eq:ovrlap}
f'_i&=  f_\tau(\vec{R}_i,\vec{P}_i')\nonumber\\
    &= \frac{1}{2/(2\pi\hbar)^3}\frac{1}{(\pi\hbar)^3}
      \sum_{k\in\tau (k\ne i)}
\,
\text{e}^{-(\vec{R}_i-\vec{R}_k)^2/2(\Delta x)^2}\\ & \quad  \times \text{e}^{-2(\Delta x/\hbar)^2(\vec{P}_i'-\vec{P}_k)^2}
\nonumber 
\end{align}
with $\tau=n$ or $p$, which estimates the probability of finding nucleons in a phase space cell of dimension $(2\pi\hbar)^3$. The factor $2/(2\pi\hbar)^3$ results from consideration of the spin in the phase-space cell.
The prefactors combine into a total factor 4.
In TuQMD, the occupation probability is calculated from an overlap of hard spheres as $f'_i=\sum_{k\in\tau (k\ne i)} (O_{ik}^{(x)}/\frac43\pi R_x^3)(O_{ik}^{(p)}/\frac43\pi R_p^3)$, where $O_{ik}^{(x)}$ ($O_{ik}^{(p)}$) is the volume of the overlap region of spheres with the radius $R_x$ ($R_p$) of nucleons $i$ and $k$ in coordinate (momentum) space.
In IQMD-IMP, the occupation probability  is $f'_i=\frac{2}{h^3}\sum_{k\in\tau (k\ne i)} O_{ik}^{(x)}O_{ik}^{(p)}$.
In CoMD, a similar procedure is used but collisions are allowed only
if the overlap $f'_i$ is less than a chosen small number (in this case, 0.08); i.e., scatterings are allowed only into essentially empty phase space cells.
On the other hand, BUU-type codes calculate the occupation probability by counting, with possible weights, numbers of test particles in a phase-space volume around the scattered particle,
but details of procedures and parameters differ among codes.  In pBUU, another procedure, which could be called an effective temperature method (see Ref.~\cite{Pawel}) is used. The distribution function in the cell around the final state of the scattered particle is fitted by a weighted sum of two deformed Fermi-Dirac distributions, and this is then used for the Pauli blocking.
Abbreviated information on the Pauli blocking treatments of the participating codes is provided in Table \ref{tab:pauli}, which in some cases updates the entries in Table III of Ref.\cite{Xu2016}.

\begin{table*}[htbp]
\newcommand{\tabincell}[2]{\begin{tabular}{@{}#1@{}}#2\end{tabular}}
\caption{
\label{tab:pauli}Pauli-blocking treatments used by different codes in the box calculation comparison.
}
\begin{tabular}{l|l|l|l}
\hline
\textbf{Code name} & \tabincell{c}{ Occupation probability $f_i$ }& \tabincell{c}{Blocking probability} \footnotemark[1] & \tabincell{c}{Additional \\ constraints} \\
\hline
\textbf{BUU-VM} &\tabincell{l}{in sphere \footnotemark[2], $\text{R}_x=2.76$ fm, $\text{R}_p=59.04$ MeV/$c$} & \tabincell{c}{$1-(1-f_i)(1-f_j)$}  & no \\
\hline
\textbf{GiBUU} & \tabincell{l}{in cube \footnotemark[2], $\Delta x=1.4$ fm, $\Delta p=68$ MeV/$c$} & \tabincell{c}{$1-(1-f_i)(1-f_j)$} & no \\
\hline
\textbf{IBUU} & \tabincell{l}{in cube \footnotemark[2], $\Delta x=2.0$ fm, $\Delta p=100$ MeV/$c$  \footnotemark[3]} & \tabincell{c}{$1-(1-f_i)(1-f_j)$ } & no \\
\hline
\textbf{pBUU} & in same and adjacent spatial cells \footnotemark[4]
& \tabincell{c}{$1-(1-f_i)(1-f_j)$}
& no \\
\hline
\textbf{RVUU} & \tabincell{l}{in cube \footnotemark[2], $\Delta x=2$ fm, $\Delta p=100$ MeV/$c$ } & \tabincell{c}{ $1-(1-f_i)(1-f_j)$ } & no\\
\hline
\textbf{SMASH} & \tabincell{l}{in sphere \footnotemark[2], $\text{R}_x=2.2$ fm, $\text{R}_p=80$ MeV/$c$  \footnotemark[5]} & \tabincell{c}{ $1-(1-f_i)(1-f_j)$ } & no \\
\hline
\textbf{SMF} & \tabincell{l}{in sphere \footnotemark[2], $\text{R}_x=2.53$ fm, $\text{R}_p=29$ MeV/$c$  \footnotemark[6]} & \tabincell{c}{$1-(1-f_i)(1-f_j)$}  & no \\
\hline
\hline
\textbf{CoMD}  & overlap of hard spheres \footnotemark[7],  & $f_i^{\prime}$, $f_j^{\prime}<f_{\mathrm{max}}=1.08$ \footnotemark[10]  & no  \\
\hline
\textbf{ImQMD} & \tabincell{l}{overlap of wavepackets \footnotemark[8], $(\Delta x)^2=2\ \text{fm}^2$ } & \tabincell{c}{$1-(1-f_i)(1-f_j)$}  & no \\
\hline
\textbf{IQMD-BNU} & \tabincell{l}{overlap of wavepackets \footnotemark[8], $(\Delta x)^2=2\ \text{fm}^2$ } & \tabincell{c}{$1-(1-f_i)(1-f_j)$} & no  \\
\hline
\textbf{IQMD-IMP}  & \tabincell{l}{overlap of hard spheres \footnotemark[9], $R_x=3.367$\ \text{fm}, $R_p=89.3$ MeV/$c$} & \tabincell{c}{$1-(1-f_i)(1-f_j)$} & no \\
\hline
\textbf{JAM} & \tabincell{l}{overlap of wavepackets \footnotemark[8], $(\Delta x)^2=2\ \text{fm}^2$ } & \tabincell{c}{$1-(1-f_i)(1-f_j)$} & no\\
\hline
\textbf{JQMD} & \tabincell{l}{overlap of wavepackets \footnotemark[8], $(\Delta x)^2=2\ \text{fm}^2$ }  & \tabincell{c}{$1-(1-f_i)(1-f_j)$}  & no \\
\hline
\textbf{TuQMD} & \tabincell{l}{overlap of hard spheres \footnotemark[11], $R_x=3.0$\ \text{fm}, $R_p=240$ MeV/$c$ } & $1-(1-f_i)(1-f_j)$ & \tabincell{c} {yes}  \\
\hline
\textbf{UrQMD} & \tabincell{l}{overlap of wavepackets \footnotemark[8], $(\Delta x)^2=2\ \text{fm}^2$ } & \tabincell{c}{$1-(1-f_i)(1-f_j)$}  & yes \footnotemark[12]\\
\hline
\footnotetext[1] {Occupation probability $f_i$ replaced by 1 if $f_i>1$.}
\footnotetext[2] {Occupation in spherical or cubic phase space cell with given dimensions.}
\footnotetext[3] {Interpolation among neighboring phase-space cells.}
\footnotetext[4] {See Ref.~\cite{Pawel} for details.}
\footnotetext[5] {With weighting in coordinate space; see ref.~\cite{Weil}.}
\footnotetext[6] {Gaussian weight in momentum space.}
\footnotetext[7] {See explanation of hard sphere overlap in the text below Eq.~\eqref{eq:ovrlap}.}
\footnotetext[8] {Overlap of wave packets, Eq.~\eqref{eq:ovrlap}, with given width $(\Delta x)^2$.}
\footnotetext[9] {$f'_i=\frac{2}{h^3}\sum_{k\in\tau (k\ne i)} O_{ik}^{(x)}O_{ik}^{(p)}$, see explanation of hard sphere overlap in the text below Eq.~\eqref{eq:ovrlap}.}
\footnotetext[10] {$f_i^{\prime}$ is the occupation of the final cell, including the scattered particle, see Ref.~\cite{Pap05} for details.}
\footnotetext[11] {$f'_i=\sum_{k\in\tau (k\ne i)} (O_{ik}^{(x)}/\frac43\pi R_x^3)(O_{ik}^{(p)}/\frac43\pi R_p^3)$ with a surface correction is applied, see Refs.~\cite{Aichelin91, Aic86} for details.}
 \footnotetext[12] {Phase-space constraint: ${{4\pi } \over 3}r_{ik}^3{{4\pi } \over 3}p_{ik}^3 \ge  {( {{h \over 2}} )^3}/4$.}

\end{tabular}
\end{table*}

\subsection{Evolution of the momentum distribution}
\label{evolution momentum}

\begin{figure*}[htpb]
\includegraphics[width=0.7\textwidth]{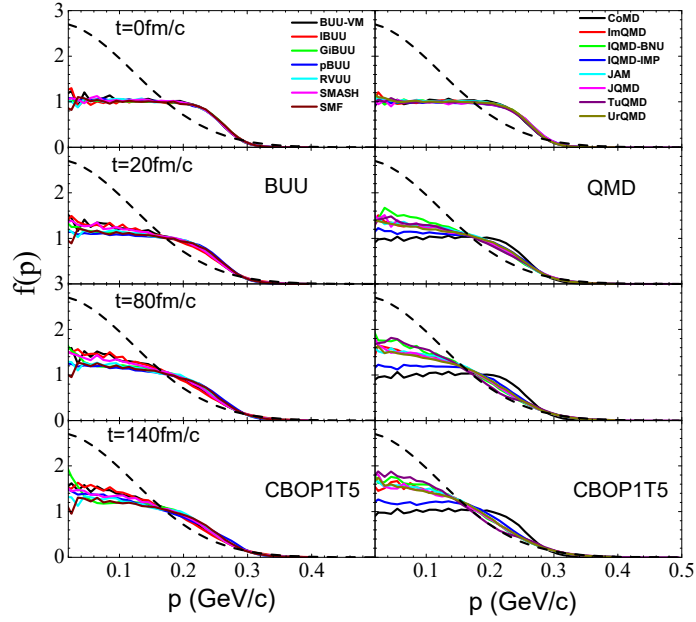}
\caption{(Color online) Momentum distributions in cascade calculations with Pauli blocking at $t$=0, 20, 80, and 140 fm/$c$ from top to bottom for $T$= 5 MeV  for BUU codes (left panels) and QMD codes (right panels) as identified in the legend. The dashed lines are ideal Boltzmann distributions at $T_B$ = 15.833 MeV.} \label{denevo_b7}
\end{figure*}

The evolution of the momentum space distribution with Pauli blocking for an initialized Fermi-Dirac distribution at $T=5$ MeV is shown in Fig.~\ref{denevo_b7}. A corresponding figure for $T=0$ MeV (mode CBOP1T0) looks very similar, except that the initial distribution is a sharp Fermi sphere. In principle, the initialized distributions should be stable in time, if the Pauli blocking were perfectly efficient. It is seen, however,  that the distributions evolve away from the Fermi-Dirac distribution toward a Boltzmann distribution.
This is the effect, mentioned above, of the coarse-graining of the phase space distribution in the sampling, effectively introducing  dissipation.
The progress of the evolution differs for different codes. Over the monitored time, the  BUU codes succeed rather well to preserve the fermionic character of the system, while QMD codes generally do worse, except for the CoMD code.
The effectiveness of blocking is different between BUU and QMD codes, and also varies within the QMD family. The different behaviors in Fig.~\ref{denevo_b7} are clearly due to the calculation of the blocking factors. In the modes CBOP2T0 and CBOP2T5, where the blocking factors are fixed from the initial Fermi-Dirac distributions, the calculated dynamic distributions (not shown here) are stable in all codes and coincide with the prescribed Fermi-Dirac distributions with good accuracy at all times.

\begin{figure*}[htpb]
\includegraphics[width=0.7\textwidth]{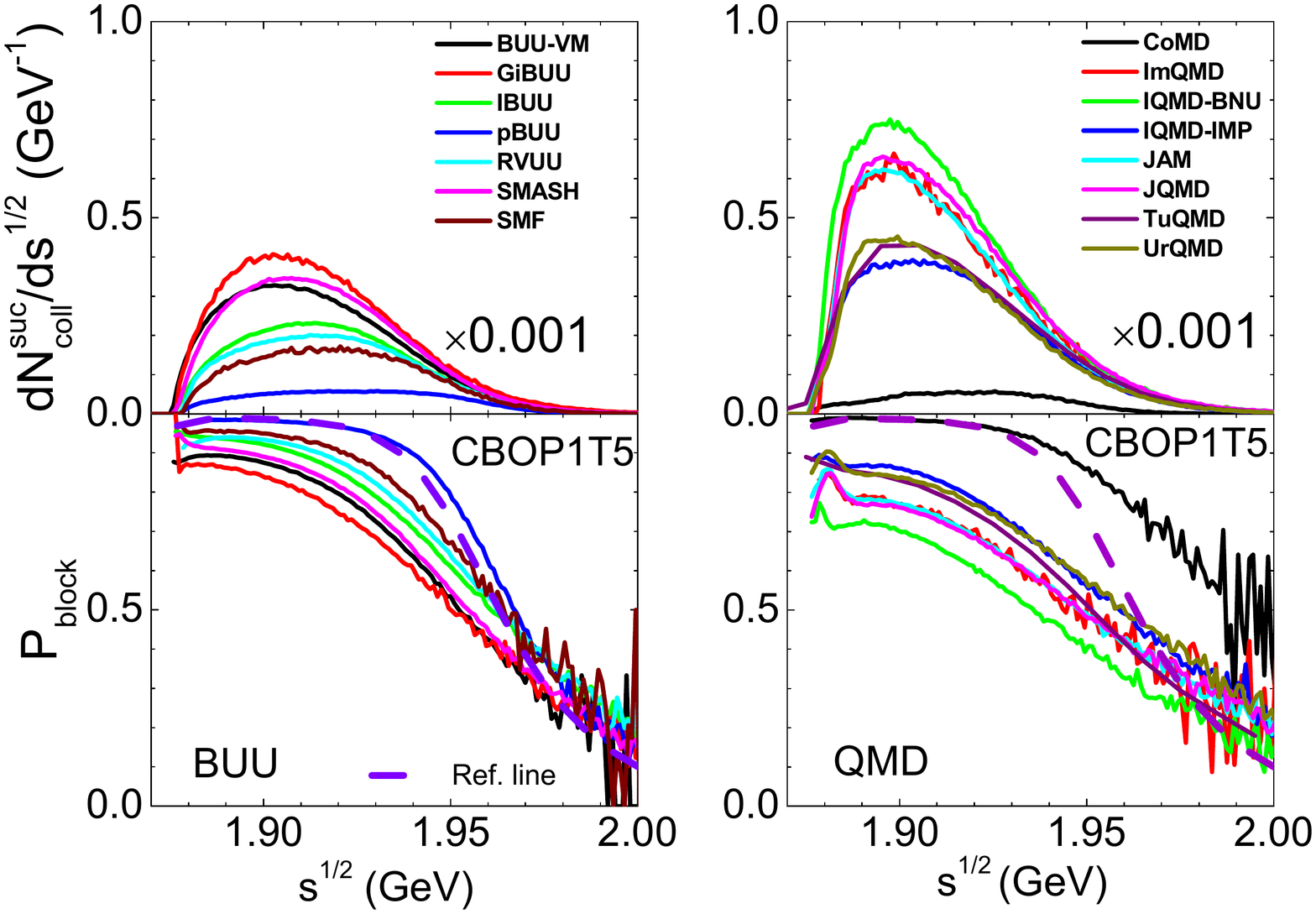}
\caption{(Color online) Top: Center-of-mass energy distribution of successful collision number averaged over time for $T$ = 5 MeV in CBOP1.  Bottom: Averaged blocking factors as a function of center-of-mass energy for colliding pairs. BUU (QMD) codes are shown on the left (right), as identified in the legend.
The thick dashed-purple lines are for the ideal Fermi-Dirac distribution at $T=5$ MeV in the relativistic case.  }
\label{collt_b7}
\end{figure*}

Differences in the Pauli blocking between different codes can further be seen in Fig.~\ref{collt_b7}, where the energy distributions of the successful collisions are shown in the upper panels, and the time averaged blocking factors, i.e., the ratio $P_{\text{block}}=\langle 1-(dN/dt)_{\text{success}}/(dN/dt)_{\text{attemp}}\rangle$ in the lower panels.
Results from the ideal Fermi-Dirac distribution at $T=5$ MeV are also shown as thick-dashed purple lines.
The energy threshold for a collision in the NN rest frame is $\sqrt{s}=1.876$ GeV.  For reference, the average center-of-mass energy for a pair in a Fermi-Dirac distribution of $T=5$ MeV is about 1.92 GeV and the average center-of-mass energy when both particles are near the Fermi surface is 1.95 GeV.
The majority of the collisions are seen to occur at lower energies close to the threshold, where the collisions should be suppressed most effectively.
Near the Fermi energy, the collision rates are seen to converge better among the codes as the blocking becomes less important.
An exception is CoMD, where the blocking probability is higher than that given by the prescribed Fermi-Dirac distribution. This is a result of the procedure mentioned above, that the final states have to be essentially empty for the collision to be allowed. On the other hand, this leads to a much better blocking for the states of lower momentum inside the Fermi sphere. Also, pBUU approaches this distribution very well, due to the effective temperature method, as mentioned
in Subsec.~\ref{phase space occupation}.

\begin{figure*}[htpb]
\includegraphics[width=0.7\textwidth]{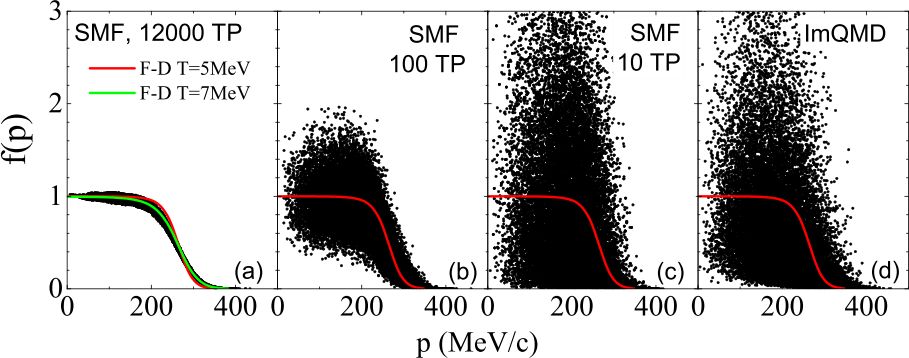}
\caption{(Color online) Scatter plots for momentum space occupation for the final states of all collisions in the first time step of the simulation (see text) for the SMF and ImQMD models initialized at $T$ = 5 MeV.  From left to right, the panels represent in sequence: (a) results from SMF with (effectively) about 12000 test particles per nucleon; (b) results from SMF with 100 test particles per nucleon and one run;   (c) results from SMF with 10 test particles per nucleon and 10 runs; and (d) results from ImQMD for 100 runs.  The red curves in the panels represent the Fermi-Dirac distribution at $T$ = 5 MeV; the green curve in the left panel is the Fermi-Dirac distribution at $T$ = 7 MeV.} \label{f-scatter}
\end{figure*}

As evidenced in Fig.~\ref{collt_b7}, the blocking differs considerably between the two families of codes, but also within the families. To better understand the origin of differences from integration over evolution, one can query about any differences when the momentum distributions are ensured to be the same.
Thus, we asked the code contributors to provide details of the Pauli blocking only for the first time step of the box simulation, when the momentum space distribution is still governed by the initialized Fermi-Dirac distribution. In particular, we collect the occupation probabilities in momentum space of the final states of the collision partners for all collisions in the first time step, and plot them as a scatter plot against the momentum of the final particle.
In Fig.~\ref{f-scatter}, we show in particular the results for SMF and ImQMD codes,  with the SMF results illustrated by those from various choices of the numbers of test particles per nucleon.

\subsection{Fluctuations of the phase-space occupation}
\label{phase space fluctuation}

The scatter plots in Fig.~\ref{f-scatter} make it evident that there is a considerable scatter of the simulated occupation probabilities around the initial $T=5 \, \text{MeV}$ Fermi-Dirac distribution, which represents the true underlying occupation probability.  The actual determination of the occupation probabilities in the dynamical evolution is through a sampling of the momentum space by different procedures as mentioned before. This leads to a considerable fluctuation of these occupation probabilities around the exact distribution, and to occupation probabilities both greater and
smaller than the true value.  There is a systematic difference between the BUU- and QMD-type codes as exemplified in the results from the SMF and ImQMD codes in Fig.~\ref{f-scatter} which are representative for the two types.  Specifically, we can see that fluctuations in the occupation probability are similar for ImQMD in Fig.~\ref{f-scatter} (d) (1 `test particle', 100 events) and for SMF in Fig.~\ref{f-scatter} (c) (10 test particles, 10 events). However, in BUU codes the fluctuation depends on the numbers of test particles used, as can be seen by comparing results from using 10 and 100 test particles per nucleon, respectively, in Fig.~\ref{f-scatter} (b) and (c) of
Fig.~\ref{f-scatter}.  The fluctuation is considerably decreased for 100 test particles.

In a box calculation, it is possible to drastically increase the effective number of test particles, by calculating the occupation probabilities not only at the location of the collision, but over the entire box, since the result should be homogenous in coordinate space. This increases the number of test particles effectively by the ratio of the volume assigned to a test particle to the volume of the box, which here was about~120, increasing the effective number of test particles per nucleon to about 12000.
The resulting scatter plot is shown in Fig.~\ref{f-scatter} (a). Now the distribution of calculated probabilities approximates far more closely the prescribed Fermi-Dirac distribution. The scatter in the probability is greatly reduced. However, it may be noted that the dynamic occupation probability is closer to a Fermi-Dirac distribution with $T=7 \, \text{MeV}$ rather than that with $T=5 \, \text{MeV}$ of the initialization.  This can be traced to the fact that a finite Gaussian shape of the test particle in momentum space is assumed in SMF with a width of 59 MeV/$c$ to obtain a smoother representation of the momentum space.
However, the initialized momentum distribution in all codes is done with the centroids of the nucleon wave packets or extended test particles, i.e., not including the folding with the width. Then the sampled momentum distribution becomes more diffuse, which, when interpreted in terms of a Fermi-Dirac distribution of finite temperature, leads to an increase of the temperature. For example, if one starts with a sharp Fermi sphere of $T=0$ MeV, then the sampled distribution of the occupation probability acquires a diffuseness, i.e., an effective temperature. In this case for an initialized Fermi-Dirac distribution of $T=5$ MeV, this increases the temperature in effect by about 2 MeV.
From Fig.~\ref{f-scatter}, it is evident that the fluctuations in BUU codes are effectively controlled by the number of test particles, i.e., by the resolution of phase space, and can be made arbitrarily small in principle. In QMD, on the other hand, the fluctuation is established once a width of the wave packets is chosen and cannot be decreased by collecting more events. We see here an immediate consequence of the philosophy of simulating heavy-ion collisions on the sampling of phase space. The homework specification essentially corresponds to the cases of Fig.~\ref{f-scatter}(b) for BUU and Fig.~\ref{f-scatter}(d) for QMD, and thus corresponds to rather different levels of fluctuations.

\begin{figure*}[htpb]
\includegraphics[width=0.8\textwidth]{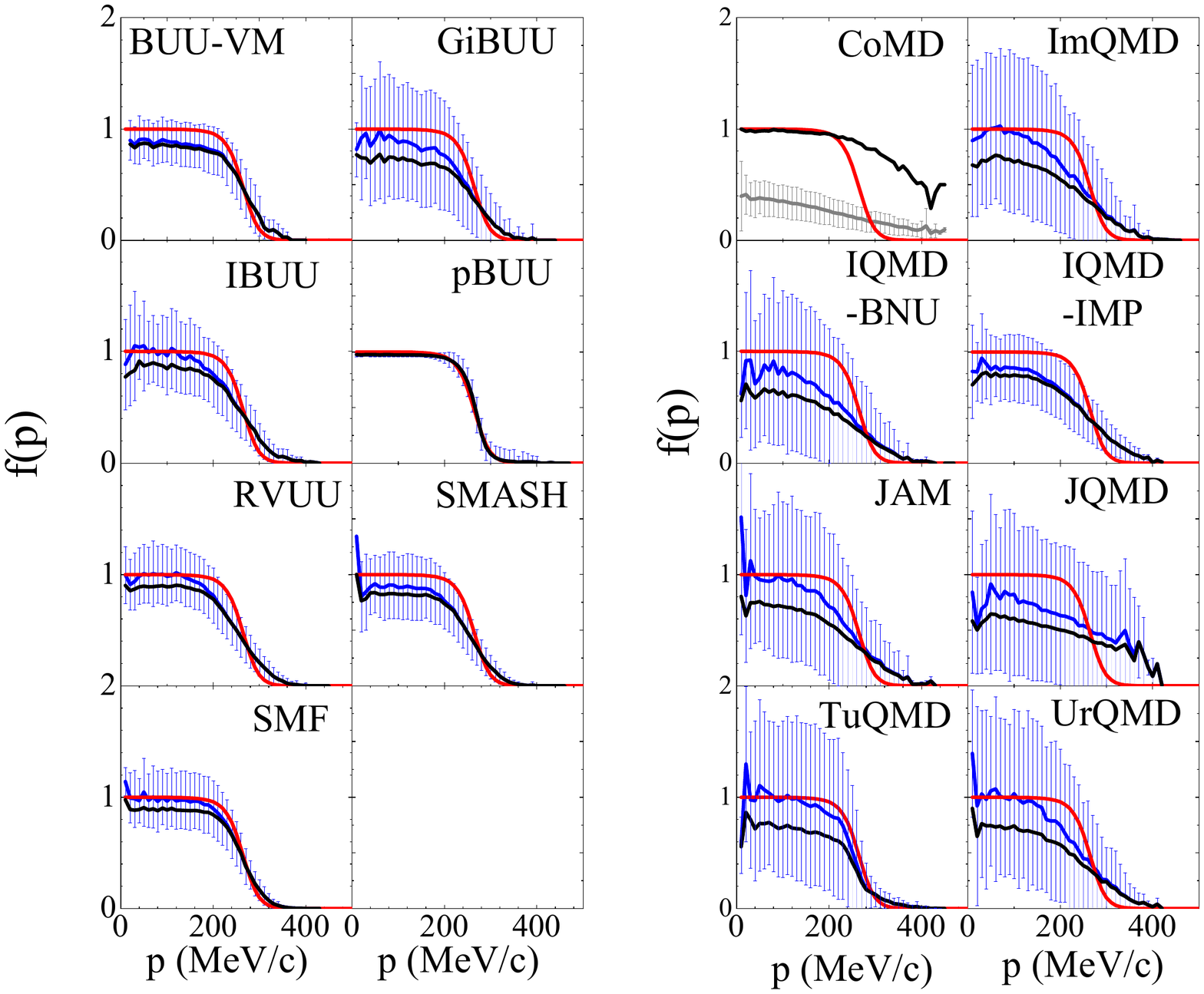}
\caption{(Color online) Distribution of occupation probabilities (blue) in the first time step of the simulation for the $T$ = 5 MeV initialization with the mean and variance shown by the blue curve and the blue error bars.  Left panels show results for BUU-type codes and right panels for QMD-type codes.  The average blocking probabilities are shown as the black curve (see text). The Fermi-Dirac distribution with $T$ = 5 MeV used for initialization is represented with the solid line (red). The gray line and error bars for CoMD are explained in the text.
}
\label{avf-pbcheck}
\end{figure*}

Figure~\ref{avf-pbcheck} gives an overview of the mean of the occupation probability and its variance obtained for the final state of all collisions in the first time step in the different BUU and QMD codes as blue lines and blue error bars. For comparison the initialized Fermi-Dirac distribution, which represents the ideal occupation probability, is plotted as a red line. It is seen that BUU codes using 100 test particles systematically give a smaller variance compared to QMD codes. But there are also differences among BUU codes, depending on the algorithm used. For example, the code pBUU with the particular effective temperature method
(see Subsec.~\ref{phase space occupation})
reproduces the given Fermi distribution almost exactly. In QMD, the variance is not only larger but the occupation distributions tend to be more smeared out, with substantial contributions for large momenta, effectively representing higher temperatures. The occupation probabilities determine the blocking of the final state, which is thus strongly influenced by these  fluctuations. The usual procedure in a case where $f>1$ is to set $f=1$, i.e., to completely block the collision. However, fluctuations to low occupation probabilities are retained. Effectively this decreases the average occupation probability, leading to overall weaker blocking than in the exact expression.
The average blocking probability shown by the black line is considerably below the red curve of the Fermi-Dirac distribution. It is generally lower for the QMD codes because of the larger fluctuation.
We recall that it is the occupation in phase space that determines the blocking. If the wave packet overlap method [see Eq.(~\ ref{eq:ovrlap})], is used, the volumes in coordinate and momentum space are inversely correlated. Then in a box calculation without inhomogeneities, the phase space volume to be sampled does not depend critically on the width parameter.   This may explain why all the QMD codes that use this method (all, except IQMD-IMP and CoMD) lead to similar fluctuations. Since these fluctuation are large also in the interior of the Fermi sphere, the momentum distributions of QMD codes in Fig.~\ref{denevo_b7} rapidly evolve  from Fermi-Dirac to Boltzmann type. It shows that the way fluctuations are introduced in QMD more quickly than in BUU destroys the fermionic character of the system to arrive at an essentially classical description.
Because of its method of blocking, the CoMD panel needs a separate explanation: The gray line and error bars represent the mean and variance of the hard sphere overlaps as a function of the momentum of a particle. Since collisions are blocked when this overlap is larger than 0.08, a blocking probability arises which is given again by the black line. It displays blocking in good agreement with the expected Fermi-Dirac curve for lower momenta, but results in overblocking for higher momentum particles, as was already seen in Fig.~\ref{collt_b7}.

\begin{figure*}[htpb]
\includegraphics[width=0.7\textwidth]{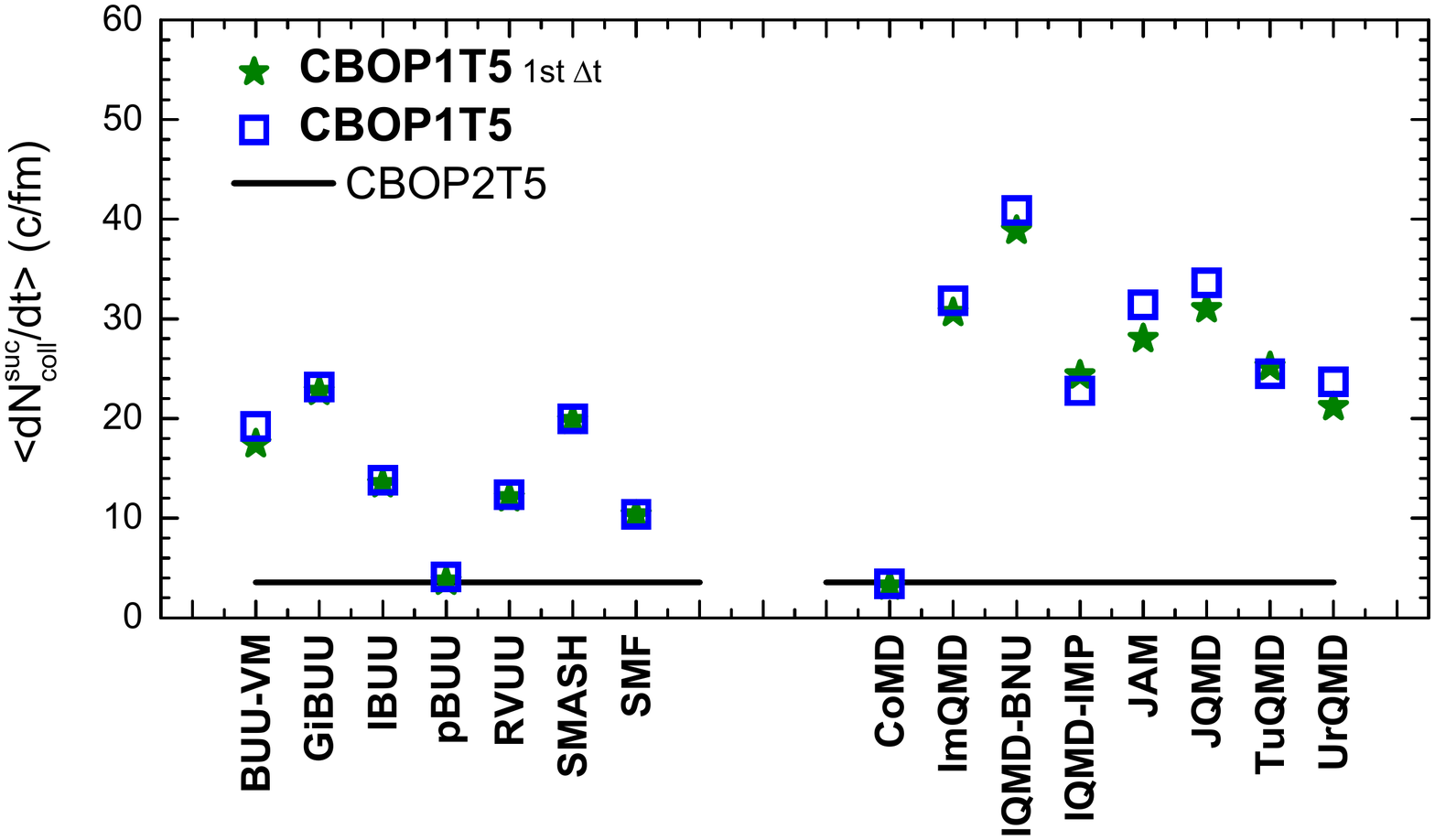}
\caption{(Color online) Successful collision rates from the different models in simulations with Pauli blocking for $T$ = 5 MeV initializations (CBOP1T5 mode). The square symbols show the results averaged over the time interval 60 - 140 fm/c, while stars are the successful collision rates for the first time step. The black line represents the reference value calculated with the basic cascade code for a fixed Fermi-Dirac blocker (CBOP2T5 mode).
}
\label{avcoll-rate-cbop1}
\end{figure*}

\subsection{Average collision rates with blocking}
\label{average rates with blocking}

Figure~\ref{avcoll-rate-cbop1} displays the successful collision rates in simulations with Pauli blocking for $T=5$ MeV initialization for the different codes (CBOP1T5 mode). Shown are the time-averaged rates for the time interval 60-140fm/$c$ as square symbols and the rates for the first time step as star symbols.
The solid (black) line represents the result of the basic cascade code, when the Pauli blocker is fixed to the initialized Fermi-Dirac distribution for $T=5$ MeV, to which the code results should be compared.
As was seen in Fig.~\ref{denevo_b7} in the first time step the momentum distributions are still the initialized Fermi-Dirac distributions, while at later times they have changed by various amounts toward Boltzmann distributions, particularly for QMD codes. Even for these codes, the effect is not very large, which might be due to the fact that the collision rates for equivalent Fermi-Dirac to Boltzmann distributions do not change very much, as was seen in Table~\ref{tab:analcoll} (though the ideal rates actually go down).
The successful collision rates for QMD are considerably higher than the BUU rates. This is consistent with the findings for the first time step in Fig.~\ref{avf-pbcheck}, that the effective blocking probability in QMD codes is lower, and is seen to be the case also for the full time evolution. Again there are considerable differences among the codes. To compare with reference values, we calculate numerically (using the options CBOP2T0 and CBOP2T5) the exact successful collision rates for a Fermi-Dirac distribution at a given temperature with the basic cascade code. For $T=0$, it should, of course, be zero.
For $T=5$ MeV, the reference successful rate value is about 3.5 $c$/fm for relativistic case (3.4 $c$/fm for non-relativistic), which is shown as a black line in Fig.~\ref{avcoll-rate-cbop1}. For $T=7$ MeV, which, as shown in Fig.~\ref{f-scatter}, approximately takes into account the effect of the finite width of the particles in momentum space, it is about 5.4 $c$/fm. The pBUU code, with the special effective temperature method to obtain blocking probabilities
(Subsec.~\ref{phase space occupation}),
succeeds rather well to obtain these limits. The SMF  rate with 100 test particles is about 10.5 $c$/fm. In the calculation with the effectively increased number of test particles, shown in Fig.~\ref{f-scatter}(a), the rate drops to about 6 $c$/fm, and thus comes close to the exact value for the effective temperature of $T=7$ MeV.
The CoMD code reproduces the reference value rather well. This may be due to the good blocking of the low momentum states, while the stronger blocking of the high momentum states, seen in Fig.~\ref{avf-pbcheck}, does not contribute so much to the total collision rate.
Generally, the collision rates of BUU codes (except pBUU) lie in the range of 10-23 $c$/fm, while for QMD codes (except CoMD) they are in the range of 23-40 $c$/fm.

\section{Discussion}
\label{discussion}

Reiterating, the evaluation of the collision term has two main steps, first determining the probability that two (test) particles collide, and second, determining whether the final states of a collision are allowed by the Pauli principle. In the box calculations these two steps can be studied separately with respect to the convergence of the different codes. Furthermore, box calculations allow one to obtain reference values from kinetic theory or from transparent numerical calculations using a basic cascade code. Thus one can make quantitative comparisons of various codes with respect to these two ingredients. Relative to the exact limits, there is an additional effect of an artificial increase of the effective temperature due to the smearing of the momenta of the (test) particles both in BUU and QMD. This effect is not very significant in view of the other differences.

\subsection{Collision probabilities}

Figure~\ref{avcoll-num} has shown that the collision probabilities are well under control in many codes, after an additional constraint was introduced
to eliminate repeated collisions between the same pair of particles which are not present in the kinetic theory. There are also good conceptual arguments to eliminate repeated collisions between the same particles, since the effective in-medium cross section is thought of as a T-matrix, which sums repeated collisions.
The residual differences between the Boltzmann theory and the simulations resulting from higher-order correlations are small enough that they are not expected to influence the interpretation of heavy-ion collisions in essential way. Most of the codes provide results that converge to the expected analytical limit to within $1\%$. For the collision probabilities, these limits provide benchmarks for all transport code. This is s especially useful for new codes under development and codes that did not participate in this comparison.

\subsection{Pauli Blocking}

The second ingredient being tested is the Pauli blocking of the final states of a collision which is important and drastically influences the evolution
of phase space occupations. Figures~\ref{avf-pbcheck} and ~\ref{avcoll-rate-cbop1} have shown that the effectiveness of the Pauli blocking, however, differs substantially among the different
codes and from the reference values. For Fermi-Dirac distributions, the successful collisions are expected to be $\langle dN_{\text{coll}}/dt \rangle
= 0$ for $T=0$ and $\sim3.5$ $c$/fm for $T=5$ MeV, respectively. However, the actual numbers are much higher, more so for QMD codes than BUU codes. Our investigation has revealed that the blocking in simulation codes is subject to large fluctuations which destroy the fermionic character of a system
in a time scale of 10-100 fm/$c$, depending on the code. Numerical fluctuations in the representation of phase space are controlled in different ways in BUU and QMD approaches. They depends on choices of parameters: in BUU on the
combination of the shape and the number of test particles, and in QMD on the chosen width of the wave packets. With the parameters chosen here, which are typical for the use in heavy-ion collisions,
the successful collision rates for BUU codes are mostly reasonably near the reference values, while those for QMD codes are consistently much above them.

\subsection{Fluctuations}

Fluctuations in transport theory are presently a much discussed issue. They are the main venue to go beyond dissipative mean field dynamics. Fluctuations have direct observable consequences in the decomposition of the dynamically unstable system into fragments and clusters during the expansion phase of a heavy-ion collision. The~amount of fluctuations is the basic difference between BUU and QMD codes.
This was discussed in greater detail in Sec. II on transport approaches.
In BUU theories, the problem is, in principle, understood: The fluctuations are related to the number of test particles per nucleon and can be reduced to zero in the limit of infinite test particle number. Beyond the statistical fluctuation, one can introduce a fluctuation term incorporated into the Boltzmann-Langevin theory. Of course, the specification of the fluctuation term is far from settled. Some codes (SMF, BLOB\cite{Napoli}, and IBL\cite{Xie13,FSZhang95}) attempt practical approximations. In QMD codes, the fluctuations are essentially controlled by the width parameter of the wave packets, which is not constrained by theory. The width parameter also controls the fluctuations due to the collision term, which relocates entire wave packets in momentum space.
Thus QMD incorporates fluctuations and correlations into the transport approach from the beginning, which, on the other hand, more quickly deteriorates the fermionic nature of the system.

Of course, there are methods to reduce the fluctuations in the calculated occupations, such as by increasing the number of test particles in BUU codes, or by using wider wave packets in QMD. However, fluctuations are physical in heavy-ion collisions and lead to observable effects. Thus they should not be arbitrarily suppressed. The question of how to properly treat the fluctuation in transport theories remains an open one.
In this comparison we do not attempt to solve this issue, but show the consequences of the different strategies and ideas in the transport codes used for the interpretation of nuclear collisions.

\subsection{Evaluation}

 As a result of the present study, we are able to make statements about the effectiveness of various procedures in the simulations of transport theories. Even though the determination of the collision probabilities converges rather well among all codes, as was seen in Fig.~\ref{avcoll-num}, there are still methods that are more effective in reproducing kinetic theory without higher order correlations. These are methods to specify collision probabilities locally by a statistical criterion, as in the mean-free-path method, rather than by geometrical criteria, as in the Bertsch prescription. This should be even more useful, if one wants to account for three-body collisions in the future.

With respect to the Pauli blocking it is more difficult to make definite statements. As discussed above, the effectiveness of Pauli blocking is strongly influenced by the amount of fluctuations in a code, the treatment of which is a matter of debate particularly between BUU and QMD approaches. But also within the same approach the effectiveness of Pauli blocking seems to be better in a local statistical description. As an example, the pBUU code, which shows the best blocking behavior, locally fits the distribution function of the final state by a weighted sum of two deformed Fermi-Dirac distributions, which is then used for the blocking probability
(Subsec.~\ref{phase space occupation}).
A particular case is the procedure adopted by CoMD, which by construction prevents occupations larger than about 1.1  of phase space cells. It was seen that this leads to very good blocking of low momentum states but to a stronger blocking at higher momenta than given by the reference values.

\section{Summary and Outlook}
\label{summary}

The present study of cascade box calculations is part of the Transport Simulation Code Evaluation Project to understand better transport simulations of heavy-ion collisions and to estimate and hopefully improve their reliability. In the previous comparison of different codes for Au + Au heavy-ion collision at intermediate energies in Ref.\cite{Xu2016}, we observed differences in observable effects, and it was suspected that these mainly originate from the treatment of the collision term. The current study suggests that while the collision probability is well under control, the effectiveness of Pauli blocking was found to differ substantially among the different codes and with respect to the reference values. This is caused by the amount of fluctuations in the representation of phase space, which is controlled in different ways in different codes. The present study allows us to make recommendations for the more effective procedures to evaluate the collision term.

In the box simulations, we discuss these
effects of the Pauli blocking in the rather extreme case
of a cold or lightly warmed Fermi-Dirac distribution. In
a real heavy-ion collision the distribution is rather quickly
characterized by high effective temperatures, where
effects of the violation of the Pauli principle should be
less important. Moreover, in many codes an option is
used to disallow collisions between particles of the same
nucleus, before there has been a collision with a particle
of the other nucleus. This also could eliminate many of
the spurious collisions seen here. However, these effects
should be checked in further tests of real heavy-ion
simulations.

We also note that the present code comparison, unlike the comparison of full heavy-ion collisions in
Ref.\cite{Xu2016}, allows us to make statements about the performance
of the codes, since the results can be compared
to the exact limits. The performance of some codes
has already been improved relative to the original codes
used in the first results of the homework. On the other
hand, one should be careful not to optimize the procedures
specifically for box calculations, which may not be
applicable in real heavy-ion collisions. For example, in a
box calculation one could improve the Pauli blocking by
averaging the occupation numbers over larger volumes of
phase space, but this would be inappropriate in heavy
ion collisions since it would average out variations which
are due to the finiteness of the system.

The code evaluation project for transport simulations will continue particularly in the box calculation mode. An ongoing project is the study of the mean field propagation.  The propagation can also be affected by fluctuations and can affect observables in reaction simulations.  In box simulations, the action of the mean field and the impact of fluctuations can be compared against limiting results from Landau theory.
A further direction is the investigation of momentum dependent mean fields, which are known to be important to give a reasonable reproduction of observed collective variables like isoscalar and isovector flow.
Another ongoing project is the study of particle production which has been important to constrain the isovector properties of the equation of state, in particular the symmetry  energy at higher densities, e.g., by observing the $\pi^-/\pi^+$ ratio.
Here predictions of simulations of heavy-ion collisions have given widely differing results. Pion observables involve new physics input for the mean fields and inelastic cross sections, but also new procedures in the simulations.
In principle, they also bring in Bose-Einstein statistics into a kinetic theory. This has recently been investigated in box calculations for gluon systems~\cite{CGreiner17}, but in our energy domain it was shown, e.g., in the pBUU code, that it is irrelevant.
We have started to test the pion production in the framework of box cascade calculations without blocking, which from this study are well under control.

\appendix
\section{Detailed comparison of attempted collisions}
\begin{figure*}
\includegraphics[width=\textwidth,page=1]{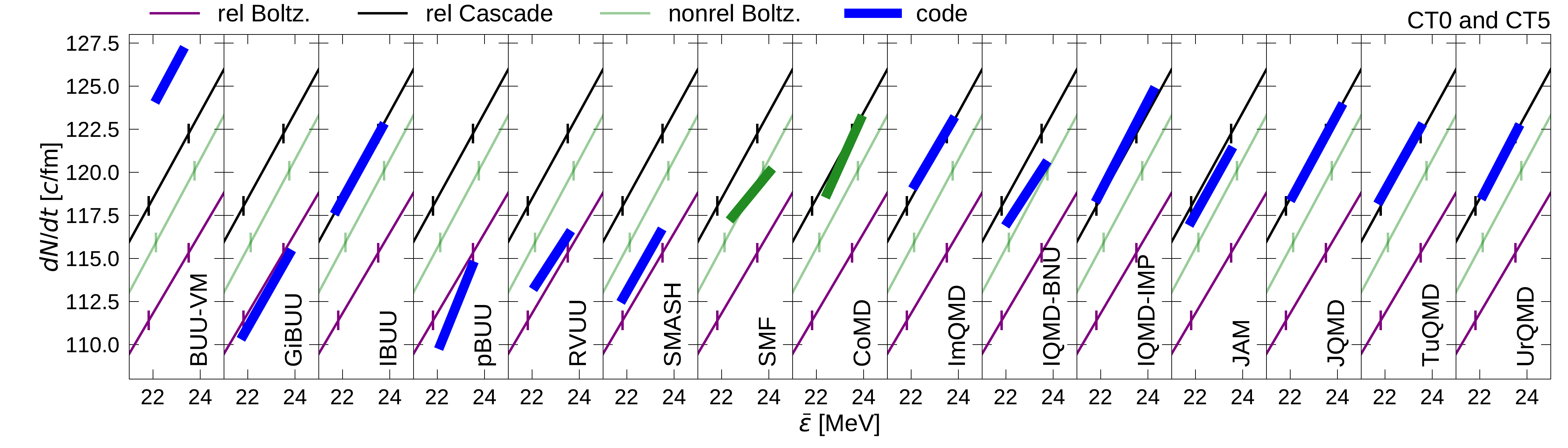}
\caption{\label{avcoll-ekin} (Color online) Collision rates in calculations without Pauli blocking vs.\ the average kinetic energy $\bar{\epsilon}$ for the different codes averaged over the time interval 60-140 fm/$c$. The thick lines (blue and green for relativistic and non-relativistic codes, respectively) connect the values for the two initial conditions of $T = 0$ and $T = 5$ MeV. The values for the reference cases from Tables \ref{tab:mut} and \ref{tab:analcoll} are given as thin lines on which the vertical bars indicate the values of $\bar{\epsilon}$ for $T=0$ and $T=5$ MeV.}

\vspace{3\baselineskip}
\includegraphics[width=\textwidth,page=2]{rateekin-1110.pdf}
\caption{\label{avcoll-ekin-cbop2} (Color online) The same as Fig.~\ref{avcoll-ekin} but for the attempted collision rates when the collisions are Pauli blocked with the probabilities evaluated with the analytic Fermi-Dirac distribution function (CBOP2T0 and CBOP2T5).  Results are not available for GiBUU.  The blue thick line for CoMD ($dN_{\text{att.}}/dt = 136.6$ and 139.0 $c$/fm for CBOP2T0 and CBOP2T5, respectively) are outside of the plotted region.}
\end{figure*}

As shown by the comparison in Sec.~\ref{results without blocking} D, the agreement of the attempted collision rates among codes and with respect to the reference values is rather satisfactory compared to the problem in Pauli blocking.  Nevertheless, if we consider the well-defined numerical procedures described in Sec.\ \ref{results without blocking} B, any deviations beyond statistical uncertainties indicate that there remain some hidden differences between codes which should preferably be eliminated in the future. The following analysis allows very precise comparison between the reference values and the code results, by separating the initialization differences discussed in Sec.~\ref{homework description} from the intrinsic differences of codes. Agreements and deviations can be analyzed more clearly here than was possible in Sec.~\ref{results without blocking}.  The comparisons here may be useful for the code authors to improve their codes.

We start with the idea that the attempted collision rate for each code should be a smooth function of the conserved total energy of the system or the average kinetic energy per nucleon denoted by $\bar{\epsilon}$.  Ideally, these functions of $\bar{\epsilon}$ should be compared between different codes and with respect to the reference values (or functions).  In our homework, the calculations were done in a relatively narrow range of $\bar{\epsilon}$ from 22 to 24 MeV (see Table \ref{tab:mut}).  Since the attempted collision rate should scale with the velocity, we may expect that the function typically behaves like $\sim\sqrt{\bar{\epsilon}}$, which can be well approximated by a linear function in the narrow range.  In our homework, the calculations have been done for two initial temperatures $T=0$ and $T=5$ MeV, corresponding to two values of $\bar{\epsilon}=\bar{\epsilon}_0$ and $\bar{\epsilon}_5$, respectively.  We therefore have sufficient information to construct the linear function for each code.  As a consequence of the initialization differences, different codes yield different values of $\bar{\epsilon}_0$ and $\bar{\epsilon}_5$.  However, the constructed linear function does not depend on the choice of the two points of $\bar{\epsilon}$ at which the function is evaluated.  Thus the comparison of the constructed linear functions is free from the influence of different choices of $\bar{\epsilon}_0$ and $\bar{\epsilon}_5$ by different codes, and therefore we can make comparisons here with much higher precision.

The results shown in Fig.~\ref{avcoll-num} are compared in a more detailed view in Fig.~\ref{avcoll-ekin}, where the collision rates are shown versus the average kinetic energy per nucleon $\bar{\epsilon}$.  For each code, the points corresponding to $\bar{\epsilon}_0$ and $\bar{\epsilon}_5$ (for the two nominal initial temperatures $T=0$ and 5 MeV, respectively) are connected by a thick line which we call the `code line' below.  We have checked that the energy is perfectly conserved in all the codes in the box simulations.
In the figure, `reference lines' are also drawn with thin lines.  Using the information on $\bar{\epsilon}_0$ and $\bar{\epsilon}_5$ in Table \ref{tab:mut} and on the reference collision rates in Table \ref{tab:analcoll}, we construct these linear functions for the different kinematic treatments.  The correct values of $\bar{\epsilon}_0$ and $\bar{\epsilon}_5$ are shown as vertical bars on the reference lines.

If a code line lies on the corresponding reference line, we can consider that the collision procedure in the code is working precisely as expected.  Good examples are JQMD and UrQMD, for which the code lines are perfectly on the reference line for the relativistic values obtained by the basic cascade code ($\delta t=\alpha\Delta t$). These two codes, as well as the basic cascade code, actually use the best relativistic time condition to judge the collision attempt, though they adopt three different formulations for the condition (see Table \ref{tab:coll}).  If a code line is shifted off the corresponding reference line but has the same slope, we consider this to be a systematic deviation.  Slight deviations in ImQMD and IQMD-IMP may be due to a different time condition (see Tables \ref{tab:coll} and \ref{tab:analcoll}), but it seems that for other codes there are also unknown or undescribed sources of systematic deviations.  The observed deviations are small in many cases except for BUU-VM.  If the slope of the code line is different from the reference line, as, e.g., for pBUU, SMF and CoMD, it is more difficult to say something about the reason, which should be investigated further.

Even if the code lines are on the same reference line (e.g., compare IBUU, JQMD, and UrQMD), we notice that they have different lengths and/or they are shifted from each other along the reference line.  This simply indicates that different codes chose different initialization parameters corresponding to different values of $\bar{\epsilon}_0$ and $\bar{\epsilon}_5$.  This problem affects the collision rates plotted in Fig.~\ref{avcoll-num}.  For example, the initialization made by JQMD corresponds to a relatively large value of $\bar{\epsilon}_5$ and therefore a high collision rate as apparently shown in Fig.~\ref{avcoll-num} for $T=5$ MeV.  By observing the situations of all the codes, the calculated collision rates in Fig.~\ref{avcoll-num} are different from those for the correct values of $\bar{\epsilon}_0$ or $\bar{\epsilon}_5$ by about 2\% at maximum.  On the other hand, this problem does not affect our comparisons here as long as we compare lines as linear functions.

To understand further the agreements and discrepancies, we examine the results from calculations of the blocking option 2 (CBOP2T0 and CBOP2T5) where the blocking probability is evaluated using the analytic Fermi-Dirac distribution function for the given temperature. No collisions are successful at $T = 0$ and collisions are quite rare at $T = 5$ MeV compared to the case of CT5 without blocking. The actual distribution of nucleon momenta is very well kept in the Fermi-Dirac form. Thus the CBOP2T0 and CBOP2T5 modes test the collision attempt procedures under the best control of the nucleon distribution and without the concern of correlations of repeated collisions.  Figure \ref{avcoll-ekin-cbop2} shows the collision attempt rates for the CBOP2T0 and CBOP2T5 modes in a similar way to Fig.~\ref{avcoll-ekin}.  We see that four QMD-type codes (JAM, JQMD, TuQMD, and UrQMD) and the pBUU and SMASH codes agree very well with the reference line for the relativistic case with the Fermi-Dirac distribution obtained by the basic cascade code with $\delta t=\alpha\Delta t$.  These codes are exactly the codes that use correct relativistic time conditions as discussed in Sec.~\ref{results without blocking} B with Table \ref{tab:coll}.  The IBUU code is unusual because it uses another time condition but agrees with the same relativistic Fermi-Dirac reference line.  It is understood that the rate is higher for ImQMD, IQMD-BNU and IQMD-IMP because these codes use other time conditions.
The slopes of lines for pBUU and SMF agree with the reference lines better here than for CT0 and CT5.  For BUU-VM and CoMD (the code line is outside the plotted range), the large differences are observed between CT0 (CT5) without blocking and CBOP2T0 (CBOP2T5) with well-defined blocking.  The GiBUU did not provide results for Fig.~\ref{avcoll-ekin-cbop2}.

In this appendix, comparisons were made for the attempted collision rates with much higher precision than in Sec.~\ref{results without blocking}.  We have clearly seen very good agreements with the expected reference values as well as some deviations, depending on the codes.  Although observed deviations are not large in most cases, these diagnostics results should be useful for the code authors to improve their codes.  Problems found in Fig.~\ref{avcoll-ekin-cbop2} should be investigated urgently since they are probably related to the fundamental prescription for collision attempts.  Problems which appear only in Fig.~\ref{avcoll-ekin} (e.g., pBUU, SMASH, SMF, and JAM), or the different directions of deviation when comparing Figs.~\ref{avcoll-ekin-cbop2} and \ref{avcoll-ekin} (e.g., BUU-VM, RVUU, CoMD, and QMD-BNU), may indicate an unexpected effect of a successful collision to subsequent collision attempts.

\acknowledgments

Y.~X.~Zhang acknowledges the supports in part by National Science Foundation of China Nos.\ 11475262, 11365004, National Key Basic Research Development Program of China under Grant No.\ 2013CB834404.
Y. J. Wang and Q.F.Li acknowledge the supports in part by National Science Foundation of China Nos. 11375062, 11505057, and 11747312, and the Zhejiang Provincial Natural Science Foundation of China (No. LY18A050002).
M. Colonna acknowledge the supports from the European Unions Horizon 2020 research and innovation programme under Grant Agreement No. 654002.
P.~Danielewicz acknowledges support from the National Science Foundation under Grant PHY-1403906.
A.~Ono acknowledges support from Japan Society for the Promotion of Science KAKENHI Grant Nos.\ 24105008 and 17K05432.
M.~B.~Tsang acknowledges the support by the US National Science Foundation Grant No. PHY-1565546, travel support from CUSTIPEN (China-US Theory Institute for Physics with Exotic Nuclei) under the US Department of Energy Grant No.\ DE-FG02-13ER42025.
H.~Wolter acknowledges support by the Cluster of Excellence {\em Origin and Structure of the Universe} of the German Research Foundation (DFG).
J.~Xu acknowledges the Major State Basic Research Development Program (973 Program) of China under Contract Nos. 2015CB856904 and 2014CB845401, the National Natural Science Foundation of China under Grant Nos.\ 11475243 and 11421505, the "100-talent plan" of Shanghai Institute of Applied Physics under Grant Nos.\ Y290061011 and Y526011011 from the Chinese Academy of Sciences, and the Shanghai Key Laboratory of Particle Physics and Cosmology under Grant No.\ 15DZ2272100.
C. M. Ko acknowledges the support by by the US Department of Energy under Contract No. DE-SC0015266 and the Welch Foundation under Grant No. A-1358. B.A. Li acknowledges the U.S. Department of Energy, Office of Science, under Award Number DE-SC0013702 and the National Natural Science Foundation of China under Grant No. 11320101004.
T. Ogawa acknowledges support from Japan Society for the Promotion of Science KAKENHI Grant  No. 26790072.
D.Oliinychenko and H.Petersen acknowledge funding of a Helmholtz Young Investigator Group VH-NG-822 from
the Helmholtz Association and GSI and support by the Helmholtz International Center for the Facility for Antiproton and Ion Research (HIC for FAIR) within the framework of the Landes-Offensive zur Entwicklung Wissenschaftlich-Oekonomischer Exzellenz (LOEWE) program launched by the State of Hesse.  D. O. was also supported by the U.S. Department of Energy, Office of Science, Office of Nuclear Physics, under contract number DE-AC02-05CH11231 and received support within the framework of theBeam Energy Scan Theory (BEST) Topical Collaboration. F.S.Zhang acknowledges National Nature Science Foundation of China under Grant No.11635003, 11025524, 11161130520.

The writing committee (consisting of the first 8 authors) would like to acknowledge the generous financial support from the director of the Facility for Rare Isotope Beams (FRIB) in hosting writing sessions in the 2017 International Collaboration in Nuclear Theory (ICNT). Writing sessions by (JX, AO, MBT and YXZ) were also hosted at the Chinese Institute of Atomic Energy (CIAE) with support from CIAE funding and State Administration of Foreign Experts Affairs, No. T170517002. The stay of AO during the ICNT meeting was supported by National Science Foundation under Grant No. PHY-1430152 (JINA Center for the Evolution of the Elements).



\begin{thebibliography}{99}
\bibitem{thanksAO} The basic cascade code is written by Akira Ono.
\end{thebibliography}
\end{document}